\documentclass[preprint,revtex4]{emulateapj}

\usepackage{apjfonts}
\usepackage{xspace}
\usepackage[colorlinks=true,linkcolor=blue,citecolor=blue]{hyperref}

\newcommand{\tdefit}{{\tt TDEFit}\xspace}
\newcommand{\Log}{{\rm Log}\xspace}
\newcommand{\V}{{\cal V}}
\newcommand{\rsep}{\leq x \leq}

\shorttitle{A UV-bright transient from ROTSE}
\shortauthors{Vink\'o et al.}

\begin{document}

\title{A Luminous, Fast Rising UV-Transient Discovered by ROTSE: a Tidal Disruption Event?}

\author{J. Vink\'o\altaffilmark{1,2}, 
F. Yuan\altaffilmark{3,4,5},
R. M. Quimby\altaffilmark{6,7},
J. C. Wheeler\altaffilmark{1},
E. Ramirez-Ruiz\altaffilmark{8},
J. Guillochon\altaffilmark{9,8},
E. Chatzopoulos\altaffilmark{1,10}, 
G. H. Marion\altaffilmark{1} and 
C. Akerlof\altaffilmark{3}
}

\email{vinko@astro.as.utexas.edu}

\altaffiltext{1}{Department of Astronomy, McDonald Observatory, University of Texas at Austin, TX, 78712, USA} 
\altaffiltext{2}{Department of Optics and Quantum Electronics, University of Szeged, D\'om t\'er 9, Szeged, 6720 Hungary}
\altaffiltext{3}{Physics Department, University of Michigan, Ann Arbor, MI 48109, USA}
\altaffiltext{4}{Research School of Astronomy and Astrophysics, Australian National University, Weston Creek ACT 2611, Australia}
\altaffiltext{5}{ARC Centre of Excellence for All-sky Astrophysics}
\altaffiltext{6}{Kavli IPMU (WPI), The University of Tokyo, 5-1-5 Kahiwanoha, Kashiwa-shi, Chiba, 277-8583, Japan}
\altaffiltext{7}{Department of Astronomy, San Diego State University, San Diego, CA 92182, USA}
\altaffiltext{8}{Department of Astronomy and Astrophysics, University of California, Santa Cruz, CA 95064, USA}
\altaffiltext{9}{Harvard-Smithsonian Center for Astrophysics, Cambridge, MA, USA}
\altaffiltext{10}{FLASH Center for Computational Science, Department of Astronomy \& Astronophysics, University of Chicago, Chicago, IL 60637 USA}


\begin{abstract}
We present follow-up observations of an optical transient (OT) discovered
by ROTSE on Jan. 21, 2009. Photometric monitoring was carried out with 
ROTSE-IIIb in the optical and {\it Swift} in the UV up to +70 days after discovery.
The light curve showed a fast rise time of $\sim 10$ days followed by a steep
decline over the next 60 days, which was much faster than that implied by $^{56}$Ni 
- $^{56}$Co radioactive decay. The SDSS DR10 database contains a faint, red object at 
the position of the OT, which appears slightly extended. 
This and other lines of evidence suggest that the OT is of extragalactic origin, 
and this faint object is likely the host galaxy. 
A sequence of optical spectra obtained with the 9.2-m Hobby-Eberly Telescope (HET) between +8 and +45 
days after discovery revealed a hot, blue continuum with no visible spectral features. A few weak
features that appeared after +30 days probably originated from the underlying host. 
Fitting synthetic templates
to the observed spectrum of the host galaxy revealed a redshift of $z = 0.19$. 
At this redshift the peak magnitude of the OT is close to $-22.5$, similar to the 
brightest super-luminous supernovae; however, the lack of identifiable spectral features 
makes the massive stellar death hypothesis less likely. A more plausible explanation appears to be the tidal disruption of a sun-like star by the central super-massive black hole. 
We argue that this transient likely belongs to a class of {\it super-Eddington} tidal disruption events.
\end{abstract}


\section{Introduction}
In the past decade untargeted (``blind'') surveys revealed the existence of 
new types of transients. A good example is the case of superluminous supernovae (SLSNe):
despite of being at least an order of magnitude brighter than ``normal'' supernovae 
\citep{galyam09, quimby11, galyam12},
SLSNe were not discovered before 2005, presumably because of the absence of their birthplaces
(low-luminosity galaxies and/or galaxy cores) in the pre-selected target lists of earlier 
transient surveys \citep{quimby11, galyam12, quimby13}. 

The Texas Supernova Search \citep[TSS]{quimbyphd} discovered the
first two SLSNe, SN~2005ap \citep{quimby07} and SN~2006gy \citep{smith06} 
that became prototypes of
two distinct subclasses within SLSNe (see \citealt{quimby13} for details on discoveries).
Its successor, the ROTSE Supernova Verification Project \citep[RSVP]{fangPhD},
continued to find SLSNe, e.g. SN~2008am \citep{manos11} or SN~2008es \citep{gezari09}.
Both surveys used most extensively the 0.45-m ROTSE-IIIb telescope at McDonald Observatory, 
Texas. Although the target fields covered mostly rich galaxy clusters closer than $D\approx 200$ 
Mpc, the majority of the discovered transients ($\approx 100$ to date) occured in significantly
more distant, background galaxies. The details of the search and detection strategies are
described in \citet{quimby12}.

In this paper we report the discovery of yet another unusual transient, detected with 
ROTSE-IIIb in the course of RSVP in 2009. The internal name of the transient was 
{\it Dougie}, but sometimes it was also designated as ROTSE3J120847.9+430121. 
Although the early light curve (LC) and the 
first spectra taken with the 9.2-m Hobby-Eberly Telescope (HET) suggested a new SLSN,
follow-up spectroscopic observations did not reveal any broad spectral features,
which is unusual even among SLSNe that sometimes show peculiar spectral evolution.
Instead, the spectra continued to show only a smooth, 
cooling continuum up to a month after discovery. At the last epochs when the transient 
was detected, narrow features due to the presumed host galaxy started to appear, 
then the transient faded below the HET detection limit. 

Subsequent spectroscopic observations with the Keck telescope
confirmed the existence of the host galaxy at redshift of $z = 0.19$. This redshift
corresponds to a distance of $D = 900$ Mpc, which, when combined with photometric
data, implies an observed absolute peak brightness of $M \approx -22.6$ mag, similar to
that of the brightest SLSNe (see Sect. 3.1). 

Here  we present a detailed  account  of the  unique  observational properties of {\it Dougie} as well as an in depth description of various model alternatives for its origin.
This paper is organized as follows. In \S 2 the photometric and spectroscopic
observations for both the transient and the host  galaxy are presented. In \S 3  four alternatives for {\it Dougie}'s origin are explored: 
a core-collapse supernova,  a NS-NS merger, a GRB jet observed off-axis and  a tidal disruption of a low-mass stellar object by the central supermassive black hole, the latter of which    
is favored by the  data.  Finally, in \S 4 we summarize our results and present  our conclusions. 

\section{Observations}\label{sec:obs}

\begin{figure}
\begin{center}
\epsscale{1.2}
\plotone{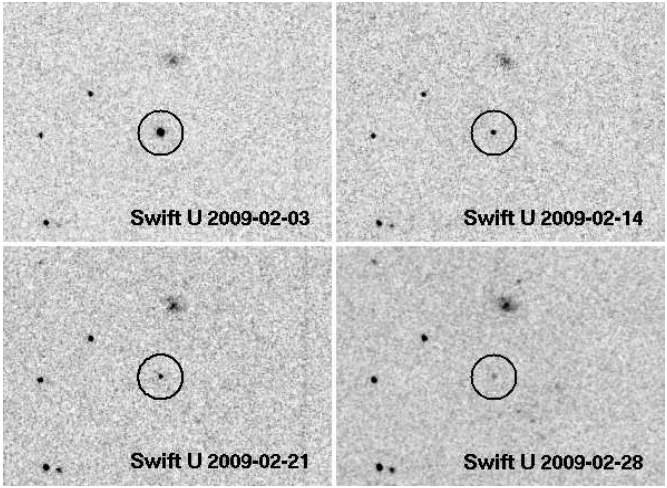}
\caption{$5 \times 4$ arcmin$^2$ fields of {\it Swift}/UVOT $u$-band frames around 
{\it Dougie} taken at different epochs. }
\label{fig:swift1}
\end{center}
\end{figure}

During its normal course of operation at McDonald Observatory, Texas, 
ROTSE-IIIb detected a new object at 
R.A.= $12^h 08^m 47.87^s$ ($\pm 0.02^s$), Dec.= $+43^o 01' 20.1"$ ($\pm 0.09"$).
The first detection occured on MJD 54852.31 (2009 Jan 21 UT 07h 26m 24s; 
UT dates are used throughout this paper), supplemented by the next detection at
UT 07h 52m 24s confirming the presence of the new object. 
At the time of discovery the apparent brightness of the transient was $\approx 17.3$ mag
(all ROTSE-IIIb unfiltered magnitudes have been converted to $R$-band magnitudes
via USNO-B1.0 and SDSS photometric calibrations, see \citealt{quimby12}). 
The ROTSE internal naming system identified the transient as 
{\it Dougie} \footnote{\tt http://www.southparkstudios.com}.

The position of the transient was checked in the SDSS DR10 catalog, and a very
faint object, SDSS J120847.77+430120.1 was found at $\approx 1.4$ arcsec distance
from {\it Dougie}. The object looks slightly more extended than nearby stars
on the combined SDSS DR10 frame, thus, the SDSS pipeline classified this object
as a galaxy and determined a photo-z = $0.207$ $\pm 0.017$ as the redshift
estimate. Our subsequent spectroscopic observation (\S 2.5) confirmed the
galaxy classification. 
We propose that this object is the host galaxy of the transient, and show
below that our measurements support the likely extragalactic origin of {\it Dougie}.

\subsection{Photometry}

\begin{figure}
\begin{center}
\plotone{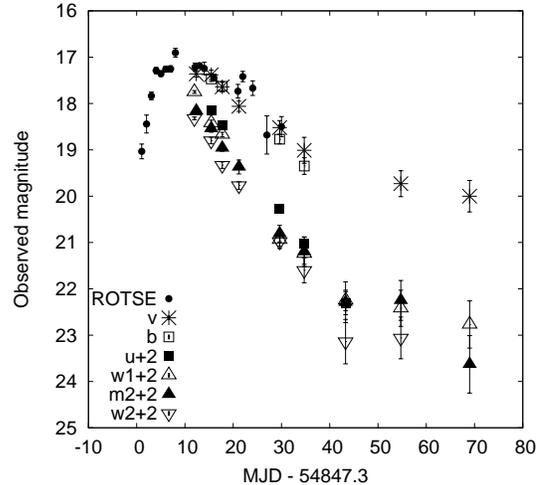}
\caption{The light curves observed with ROTSE and {\it Swift}/UVOT. The UV data has been shifted
down by 2 mag to enable comparison. }
\label{fig:lc}
\end{center}
\end{figure}

\begin{figure}
\begin{center}
\epsscale{.7}
\plotone{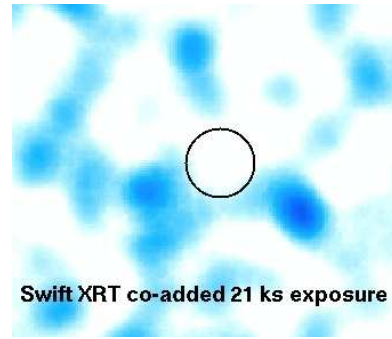}
\caption{{\it Swift}/XRT co-added 21 ks exposure of the field around the 
expected position of {\it Dougie}. The field-of-view
is the same as in Fig.~\ref{fig:swift1}. No source is detected at the position 
of the transient.} 
\label{fig:swift2}
\end{center}
\end{figure}

Tracing back in the ROTSE observational archive, the earliest detection of  
{\it Dougie} was found on the frames obtained at 4 days before discovery
(MJD 54848.34, 2009 Jan 17) when the OT was at $\approx 19.0$ mag. 
The last pre-discovery non-detection (limiting magnitude $\approx 19.6$ mag) 
occurred on 2009 Jan 15 (MJD 54846.3), 
6 days before discovery. In the following we assume that the outburst
started during the 2 days between the last non-detection and the first successful
detection, and set the ``moment of first light'' as $t_0$ = MJD 54847.3 $\pm 1.0$
(hereafter we use the term ``first light'' to refer to 
the first observable appearance of the transient, thought to be the
moment of shock breakout in SNe, for example).

Photometric follow-up observations with ROTSE-IIIb started immediately after discovery,
and continued up to 2009 Feb 15 when {\it Dougie} was at $\approx18.5$ mag. 

Additional photometric data were collected by {\it Swift}/UVOT in 3 optical 
($u$,~$b$,~$v$) and 3 ultraviolet (UV) filters ($uvw1$,~$uvm2$,~$uvw2$) after
triggering {\it Swift} in Target-of-Opportunity (ToO) observing mode. 
The UVOT observations started on 2009 Jan 28 and continued up to 2009 March 26
when the transient was below or close to the detection limit of UVOT in all filters.
Fig.~\ref{fig:swift1} illustrates the temporal evolution of the OT 
on UVOT $u$-band frames.

Photometry of {\it Dougie} was computed applying aperture photometry on the 
{\it Swift}/UVOT Level-2 (sky) frames, using the calibration by \citet{poole08}. 

The photometry on the ROTSE-IIIb frames was performed by PSF-fitting on the 
template-subtracted frames. The results were converted to $R$-band magnitudes
as noted above. All photometric data are collected in Table~\ref{tab:rotse}
and \ref{tab:uvot}. The light curves are shown in Fig.~\ref{fig:lc}.

The redshift of  {\it Dougie}'s host galaxy was estimated spectroscopically (see \S 2.3)
as $z = 0.191$, which corresponds to a luminosity distance of 
$D_L = 897$ Mpc assuming $\Lambda$-CDM cosmology with $H_0 = 73$ 
km~s$^{-1}$~Mpc$^{-1}$. Using this distance, the observed peak 
$R$-band magnitude ($\approx 17.2$ mag) translates to -22.6 mag absolute. 
As noted above, this peak brightness is comparable to that
of the most luminous SLSNe. 

\begin{table}
\begin{center}
\caption{\label{tab:rotse} ROTSE-IIIb photometry of {\it Dougie}}
\begin{tabular}{cccc}
\hline
\hline
MJD & $R$ & error$^{\rm a}$ & $3 \sigma$ limit \\
 & (mag) & (mag) & (mag) \\
\hline
54848.34 &19.03 & 0.15 & 19.90 \\
54849.34 &18.44 & 0.19 & 19.45 \\
54850.33 &17.83 & 0.08 & 19.59 \\ 
54851.32 &17.28 & 0.06 & 19.98 \\
54852.32 &17.35 & 0.06 & 20.23 \\
54853.32 &17.25 & 0.06 & 19.35 \\
54854.32 &17.25 & 0.06 & 19.18 \\
54855.31 &16.90 & 0.09 & 18.49 \\
54859.38 &17.21 & 0.08 & 19.30 \\
54860.32 &17.18 & 0.04 & 19.87 \\
54861.29 &17.24 & 0.13 & 19.04 \\
54863.29 &17.44 & 0.06 & 19.86 \\
54868.28 &17.73 & 0.14 & 18.87 \\
54869.27 &17.41 & 0.11 & 18.47 \\
54871.33 &17.66 & 0.15 & 18.25 \\
54874.27 &18.67 & 0.41 & 18.76 \\
54877.30 &18.49 & 0.21 & 18.82 \\
\hline
\hline
\end{tabular}
\end{center}
$^{\rm a}$without the $\approx 0.1$ mag zero-point uncertainty
\end{table}

\subsection{X-ray observations}\label{sec:xray}

Simultaneously with the {\it Swift}/UVOT observations, {\it Dougie} was monitored
by {\it Swift}/XRT in X-rays between 0.2 and 10 keV. A total 
of 21 ks observations were collected, extracted and added up using 
the appropriate tools in HEAsoft. 

Fig.~\ref{fig:swift2} shows a $5 \times 4$ arcmin field of the co-added XRT frame
(after applying gaussian smoothing) centered on the position of {\it Dougie}. 
No source is detected at the position of the transient. Using 
WebPIMMS\footnote{{\tt http://swift.gsfc.nasa.gov/Tools/w3pimms.html}}, the $3 \sigma$
detection limit, after correcting for the Galactic hydrogen column
density of $N_H = 1.28 \times 10^{20}$ cm$^{-2}$ \citep{kalbe} and 
assuming $\gamma=1$ for the photon index, was found to be
$f_X(3 \sigma) = 6.88 \times 10^{-14}$ erg s$^{-1}$ cm$^{-2}$, corresponding to 
$L_X < 6.6 \times10^{42}$ erg s$^{-1}$ for the upper limit of {\it Dougie}'s 
X-ray luminosity.

\begin{table}
\begin{center}
\caption{\label{tab:uvot} {\it Swift}/UVOT photometry of {\it Dougie}}
\begin{tabular}{cccc}
\hline
\hline
MJD & filter & $m^a$ & error$^b$ \\
    &  & (mag) & (mag) \\
\hline
54859.65 & v & 17.36 & 0.07 \\
54862.74 & v & 17.38 & 0.10 \\
54865.02 & v & 17.64 & 0.11 \\
54868.49 & v & 18.06 & 0.11 \\
54876.92 & v & 18.52 & 0.15 \\
54882.02 & v & 19.01 & 0.28 \\
54902.03 & v & 19.73 & 0.28 \\
54916.28 & v & 20.00 & 0.34 \\
\hline
54862.74 & b & 17.47 & 0.05 \\
54865.03 & b & 17.63 & 0.09 \\
54876.93 & b & 18.77 & 0.10 \\
54882.03 & b & 19.35 & 0.18 \\
\hline
54862.74 & u & 16.15 & 0.04 \\
54865.03 & u & 16.46 & 0.05 \\
54876.93 & u & 18.27 & 0.08 \\
54882.03 & u & 19.02 & 0.14 \\
54890.59 & u & 20.31 & 0.25 \\
\hline
54859.26 & uvw1 & 15.75 & 0.03 \\
54862.74 & uvw1 & 16.42 & 0.06 \\
54865.03 & uvw1 & 16.67 & 0.05 \\
54876.92 & uvw1 & 18.94 & 0.14 \\
54882.02 & uvw1 & 19.24 & 0.23 \\
54890.59 & uvw1 & 20.25 & 0.22 \\
54902.03 & uvw1 & 20.42 & 0.39 \\
54916.29 & uvw1 & 20.77 & 0.51 \\
\hline
54859.65 & uvm2 & 16.16 & 0.04 \\
54862.74 & uvm2 & 16.54 & 0.08 \\
54865.03 & uvm2 & 16.96 & 0.07 \\
54868.49 & uvm2 & 17.37 & 0.15 \\
54876.92 & uvm2 & 18.81 & 0.18 \\
54882.02 & uvm2 & 19.20 & 0.32 \\
54890.58 & uvm2 & 20.29 & 0.44 \\
54902.03 & uvm2 & 20.25 & 0.43 \\
54916.29 & uvm2 & 21.63 & 0.62 \\
\hline
54859.26 & uvw2 & 16.32 & 0.03 \\
54862.73 & uvw2 & 16.80 & 0.07 \\
54865.02 & uvw2 & 17.33 & 0.07 \\
54868.48 & uvw2 & 17.77 & 0.08 \\
54876.92 & uvw2 & 18.98 & 0.15 \\
54882.02 & uvw2 & 19.60 & 0.27 \\
54890.58 & uvw2 & 21.14 & 0.48 \\
54902.03 & uvw2 & 21.06 & 0.45 \\ 
\hline
\hline
\end{tabular}
\end{center}
$^{\rm a}$not corrected for host galaxy contamination\\
$^{\rm b}$statistical uncertainty only
\end{table}

\subsection{Spectroscopy}

\begin{figure}
\begin{center}
\epsscale{1.2}
\plotone{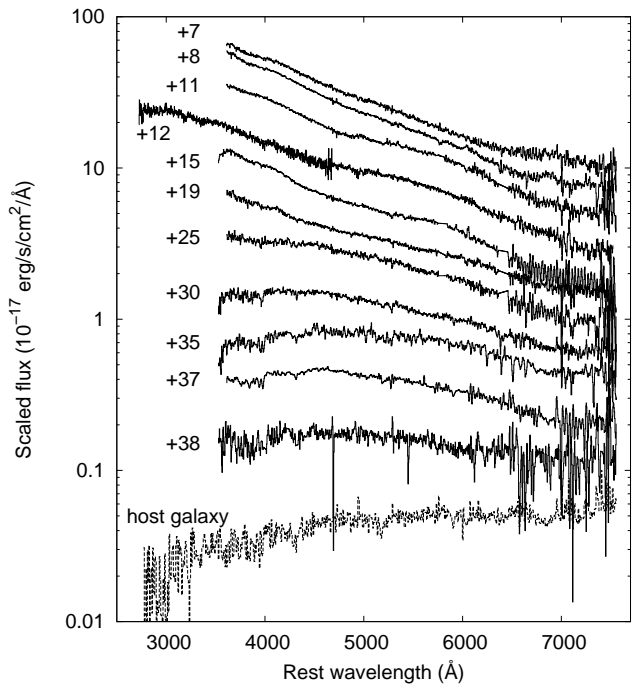}
\caption{Sequence of observed optical spectra, corrected for redshift
($z=0.191$, see text), and shifted vertically for better visibility. 
Rest-frame phases since the estimated moment of outburst (see Table~\ref{tab:hetlog})
are indicated by the labels next to each spectrum. The last spectrum is that of the
host galaxy (\S 2.5). All spectra are dominated by a cooling continuum,
and do not show any obvious spectral feature.}
\label{fig:sp}
\end{center}
\end{figure}

Optical spectra were obtained with the Marcario Low-Resolution 
Spectrograph (LRS, \citealt{hill98}) mounted on the 9.2m Hobby-Eberly
Telescope (HET, \citealt{ramsey98}) at McDonald Observatory, Texas. 
Ten spectra were collected between 2009 Jan 24 and March 02, starting
around maximum light and extending up to about 1 month thereafter.
In addition, a  spectrum was taken with the Double Spectrograph
(DBSP, \citealt{oke-dbsp}) operating on the Palomar 200-inch telescope, on 2009 Jan 30. 
The log of the spectral observations is presented in Table~\ref{tab:hetlog}. 

\begin{table*}
\begin{center}
\caption{\label{tab:hetlog} Log of spectroscopic observations}
\begin{tabular}{ccccccccl}
\hline
\hline
Date & MJD & Phase$^a$ & Exposure & Airmass & Range & FWHM & S/N$^b$ & Instr. \\  
   &  & (days) & (s) &  & (\AA) & (\AA) &  & \\
\hline
2009-01-24 & 54855.35 & +7 & 1800 & 1.16 & 4300 -- 10,000 & 19 & 42 & HET/LRS \\
2009-01-25 & 54856.32 & +8 & 1800 & 1.29 & 4300 -- 10,000 & 19 & 75 & HET/LRS\\
2009-01-29 & 54860.33 & +11 & 1800 & 1.20 & 4300 -- 10,000 & 19 & 75 & HET/LRS \\
2009-01-30 & 54861.50 & +12 & 600 & 1.04 & 3240 -- 8950 & 15 & 40 & P200/DBSP\\
2009-02-03 & 54865.30 & +15 & 1800 & 1.24 & 4300 -- 10,000 & 19 & 66 & HET/LRS\\
2009-02-07 & 54869.52 & +19 & 1800 & 1.19 & 4300 -- 10,000 & 19 & 44 & HET/LRS\\
2009-02-15 & 54877.26 & +25 & 1800 & 1.27 & 4300 -- 9000 & 19 & 32 & HET/LRS\\
2009-02-20 & 54882.50 & +30 & 1800 & 1.24 & 4300 -- 10,000 & 19 & 39 & HET/LRS\\
2009-02-26 & 54888.46 & +35 & 1800 & 1.16 & 4250 -- 10,000 & 19 & 15 & HET/LRS\\
2009-03-01 & 54891.23 & +37 & 5100 & 1.15 & 4300 -- 10,000 & 19 & 43 & HET/LRS\\
2009-03-02 & 54892.22 & +38 & 3600 & 1.31 & 4200 -- 9100 & 19 & 14 & HET/LRS\\
2009-11-11 & 55146.60 & +251 & 900/765 & 1.63 & 3300 -- 10,000 & 6 & 11 & Keck-I/LRIS\\
\hline
\hline
\end{tabular}
\end{center}
$^{\rm a}$rest-frame days since outburst assuming $T_0 = 54847.3$ MJD and $z=0.191$\\
$^{\rm b}$signal-to-noise measured at 6000 \AA
\end{table*}

All spectra were reduced in the standard way using 
IRAF\footnote{IRAF is distributed by the National Optical Astronomy Observatories,
which are operated by the Association of Universities for Research
in Astronomy, Inc., under cooperative agreement with the National
Science Foundation.}. Wavelength calibration was done based on combined
exposures of Cd and Ne spectral lamps. Flux calibration was computed
using spectra of spectro-photometric, flux standard stars taken on the same nights
when the transient was observed, which provided reliable relative fluxes
for the object's spectra. Absolute flux levels were determined by matching
the HET spectra with the flux densities from contemporaneous 
{\it Swift}/UVOT $b$ and $v$ observations. 
 
The observed spectral sequence is plotted in Fig.~\ref{fig:sp}, where the individual
spectra have been shifted vertically for clarity, and also Doppler-corrected
back to the host galaxy's  rest frame assuming $z=0.191$ (see below). 

The spectra are dominated by a smooth, hot continuum without any obviously
noticeable spectral feature. At later phases (after Feb 20) weak narrow features
appeared between 4000 -- 6000 \AA\ rest-frame wavelengths, which  are
probably due to contamination from the host galaxy (see \S 2.3). 

The combined optical + UV spectral energy distribution (SED) of {\it Dougie} was
constructed by combining the HET spectra with the {\it Swift}/UVOT photometric 
flux densities taken close to the spectroscopic observations. 
These SEDs were then corrected for Milky Way extinction using 
$E(B-V)_{\rm gal} = 0.0136$ \citep{sfd}. Reddening within the host galaxy was ignored
because of the lack of information on this parameter, but the very blue observed color of the 
transient during the early phases argues against significant in-host extinction. Finally, 
the flux contribution from the host galaxy was also subtracted from the combined UV-optical
SEDs. This correction was negligible during the early phases, but increased considerably
when the transient evolved after maximum.

\begin{figure}
\begin{center}
\plotone{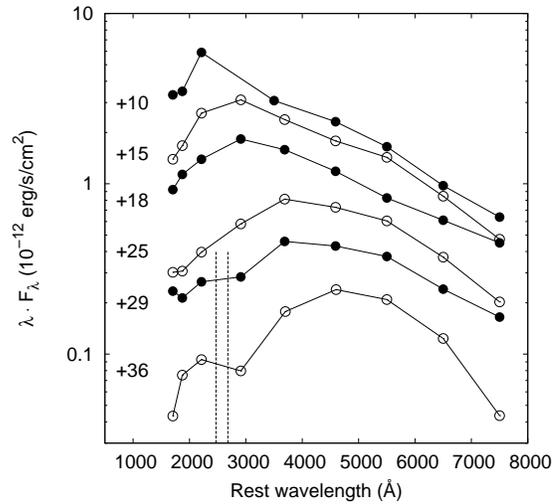}
\caption{Evolution of the optical + UV SED after correcting for host galaxy contamination. 
The phase of the transient (the elapsed time since outburst in rest-frame days) is indicated at
the left-hand side of each SED. No vertical shifts have been applied between the curves. The dotted
vertical lines mark the position of the 2470 \AA\ and the 2680 \AA\ features observed in
PS1-10bzj \citep{lunnan13} and PS1-11af \citep{chorn13}.}
\label{fig:sed}
\end{center}
\end{figure}

Fig.~\ref{fig:sed} plots the temporal  evolution of the SED  in  rest-frame days. The SED peak is observed to gradually shift from 2200 \AA\ at +10 d to 4600 \AA\
at +36 d, corresponding to $\approx 13,000$ K and $\approx 6300$ K Wien-temperatures,
respectively. The flux depression between 2500 - 3000 \AA\ appearing
after +25 d might be due to the broad UV-features observed in the spectra
of the SLSN PS1-10bzj \citep{lunnan13} and the TDE candidate PS1-11af
\citep{chorn13}; however, the resolution provided by the broadband
{\it Swift} UV filters are not adequate to unambiguously identify these features.
Alternatively, the ``UV-bump'' appearing on +36 d 
might be caused by the red leak of the {\it Swift} UV-filters.

The SEDs in Fig.~\ref{fig:sed} cannot be described by a series of single-temperature 
blackbodies: the optical continuum may suggest a higher temperature, but
the increasing flux decline in the UV is inconsistent with  the hot blackbody assumption.
It is possible that the UV is affected by strong blending due to
ionized metal lines, as usual in SNe; however, without having a well-resolved 
UV spectrum, such a conclusion cannot be proven unambiguously.

\subsection{Comparison with spectra of SLSNe}

\begin{figure*}
\begin{center}
\epsscale{1.1}
\plottwo{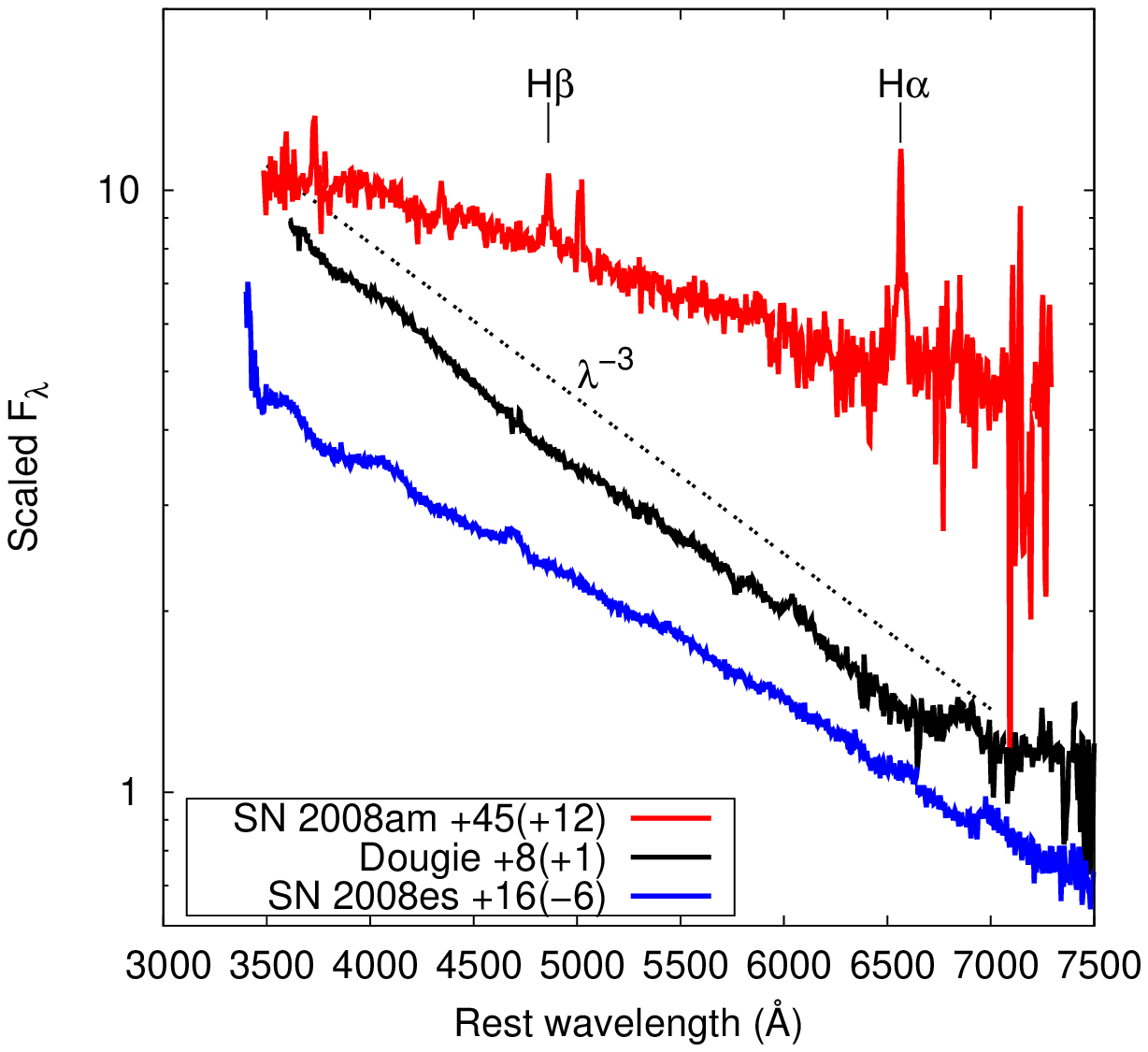}{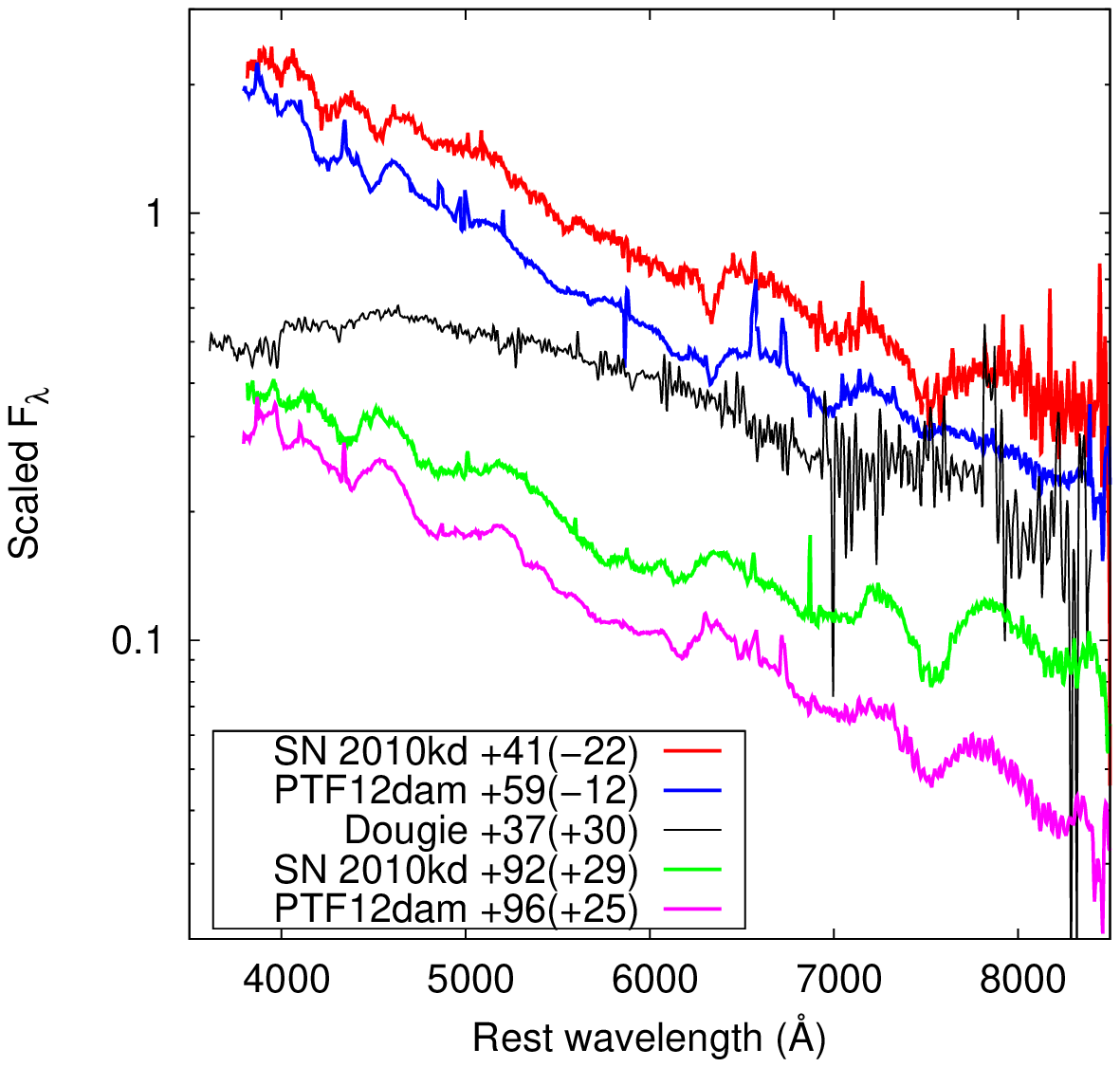}
\caption{Left: comparison of an early-phase spectrum of {\it Dougie} (black curve) 
with those of two other H-rich SLSNe (red and blue curves). 
The spectra have been scaled vertically to facilitate comparison. 
The first numbers in the legend indicate the rest-frame phase since first light, while the 
numbers in parentheses show the rest-frame days measured from the peak of the light curve.
The spectrum of {\it Dougie} appears hotter than the two other SLSNe, and its optical
continuum is approximately a power-law, $F_{\lambda} \propto \lambda^{-3}$ (dashed line; see
text for explanation). Right: one of the last observed spectra of {\it Dougie} plotted together 
with two other H-poor SLSNe observed with HET at similar phases. }
\label{fig:sne}
\end{center}
\end{figure*}

The earliest spectra of {\it Dougie} appeared similar to those of some 
SLSNe observed with HET, showing mostly a hot, featureless continuum.
This is illustrated in the left panel of Fig.~\ref{fig:sne}, where the Jan 25 spectrum
(+8 day phase after first light in rest-frame) is plotted together with the early-phase
HET spectra of two H-rich SLSNe: SN~2008am \citep{manos11} and SN~2008es
\citep{gezari09}. It is clearly observed that, unlike SN~2008am, 
{\it Dougie} did not show either hydrogen- or any other spectral
features. SN~2008es was similarly absent of features in the early spectra, 
but its late-time spectra (not shown here) contained strong, unambiguous SN
features including Balmer-lines \citep{gezari09, miller09}. 
On the contrary, as seen in Fig.~\ref{fig:sp}, none of  {\it Dougie}'s observed 
spectra show SN-like features.

Fig.~\ref{fig:sne} also illustrates  that the continuum slope of  {\it Dougie}'s early-phase
 optical spectra is  relatively well described by  a power law with $F_{\lambda} \propto \lambda^{-3}$. Pure, hotter  blackbody  spectra are  incompatible with the UV SED. 
The observed spectra  of  {\it Dougie} cannot be accurately modeled  
by either an evolving  single-temperature blackbody 
or by a  power-law spectrum  with a fixed  slope.

The right panel of Fig.~\ref{fig:sne} shows a similar comparison between the
last observed spectrum of {\it Dougie} (+37 days after first light, or
+30 days after the peak of the light curve, both measured in rest-frame) 
and spectra of SN~2010kd (Vinko et al. in prep.), 
and PTF12dam \citep{nicholl13, chen14}. Contrary to the early-phase spectrum in 
the left panel, this late-phase spectrum of {\it Dougie} is redder than 
the other SLSNe at similar post-peak phases. 
This and the lack of the spectral features make the
spectral evolution of {\it Dougie} being quite different from both H-rich and H-free SLSNe. 

We cannot rule out that the lack of broad SN features in {\it Dougie}'s  spectra 
might be simply due to an observational effect.  The broad SN features might have
appeared only at later times  when the transient faded below the HET
detection limit. Although this scenario cannot be excluded, this seems 
improbable if {\it Dougie} is thought to be similar to other fast-evolving 
SLSNe, like SN~2008es, which occurred  at a similar distance. In SN~2008es 
the broad SN features started to appear after +20 rest-frame
days \citep{miller09}, while in {\it Dougie} they failed to appear for at least
+38 rest-frame days.

Similar hot, featureless spectra have also been observed in more recent
SLSNe such as  PS1-10bzj \citep{lunnan13} and CSS121015 \citep{benetti13};
but, again, there are important differences  in the observed spectral evolution
between these SLSNe and {\it Dougie}. CSS121015 was a slowly evolving H-rich SLSN 
(its light curve peaking at +40 d rest-frame), which developed
broad H, CaII, MgII and FeII  features after +100 d \citep{benetti13}.
By contrast, PS1-10bzj was a H-poor SLSN showing rapid evolution, and by +16 d
rest frame it had also developed the usual, broad features common to all H-poor 
SLSNe \citep{lunnan13}.
Although SLSNe show some degree of diversity in their observed properties,
the lack of any broad spectral feature in  {\it Dougie}'s spectra  during the entire observable window 
is unprecedented to date.

\subsection{The host galaxy}

\begin{figure}
\begin{center}
\epsscale{1.1}
\plotone{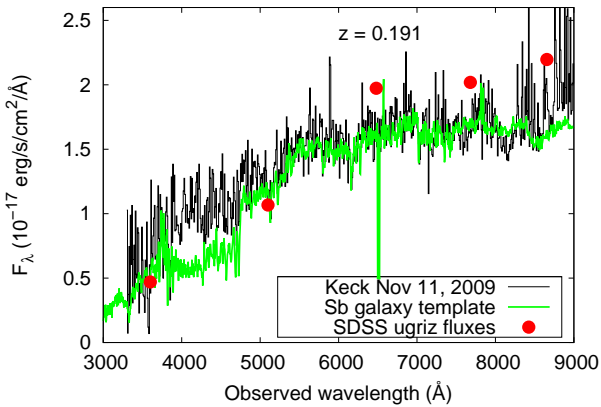}
\caption{The observed Keck spectrum of the host (black curve) compared with an
Sb-type galaxy template (green curve) redshifted to $z=0.191$. The {\it ugriz} fluxes
(red filled circles) from SDSS DR10 are also overplotted.  The host spectrum looks being 
dominated by older stellar population and does not show any sign of 
either enhanced star formation or AGN activity.}
\label{fig:host}
\end{center}
\end{figure}

The candidate  host  for {\it Dougie} is the galaxy SDSS J120847.77+430120.1.
The SDSS $ugriz$ PSF AB-magnitudes for this object are 
$u'=23.096$ ($\pm 0.428$), $g'=21.486$ (0.052), $r'=20.299$ (0.025), $i'=19.882$ (0.036),
$z'=19.510$ (0.059), while its photo-z is $0.207$ $\pm 0.017$ according to the SDSS DR10
database. 

There is no detected object in the GALEX database\footnote{\tt http://galex.stsci.edu/GR6/}
closer than 10 arcsec to this position. Adopting $m_{\rm AB} = 20.5$ mag as the limiting
magnitude for the GALEX all-sky survey, the background-corrected flux upper limit for
the host is $\approx  1.3 \times 10^{-16}$ erg s$^{-1}$ cm$^{-2}$ \AA$^{-1}$ in both 
the NUV ($\lambda 2271$ \AA) and the FUV ($\lambda 1528$ \AA) bands.  The lack of UV detection
is consistent with the photometric and spectroscopic optical observations  (see below). 
Also, there is no known X-ray or radio source in the vicinity of  {\it Dougie}'s position. According
to the SIMBAD\footnote{\tt http://simbad.u-strasbg.fr} database, the
closest radio source (WN~J1208+4301) is $\sim 1$ arcmin away and is not related
to the host. 

We have observed the candidate host galaxy  with the double-channel
Low Resolution Imaging Spectrometer (LRIS) \citep{oke-lris} 
mounted on the Keck-I telescope on 2009 Nov 11.6 UT (MJD 55146.60).
The spectrum is plotted together
with the broad-band SDSS fluxes in Fig.~\ref{fig:host}. A Sb-type galaxy template 
taken from \citet{kinney96}  is also shown for comparison.

Cross-correlation between the observed and the template galaxy spectra
revealed $z=0.191$ $\pm 0.022$ as the optimum estimate for the redshift of the host, 
which is adopted for this paper. 
This spectroscopic redshift is consistent with the photo-z estimate derived  from  the SDSS photometry. 

Note that there is no indication for either the presence of an
active galactic nucleus (AGN), or any ongoing star formation in the
host galaxy spectrum. These would produce strong, unambiguous, narrow 
emission lines that are not observed  in the  galaxy spectrum. 

\subsection{Position within the host galaxy}

\begin{figure}
\begin{center}
\epsscale{1.2}
\plotone{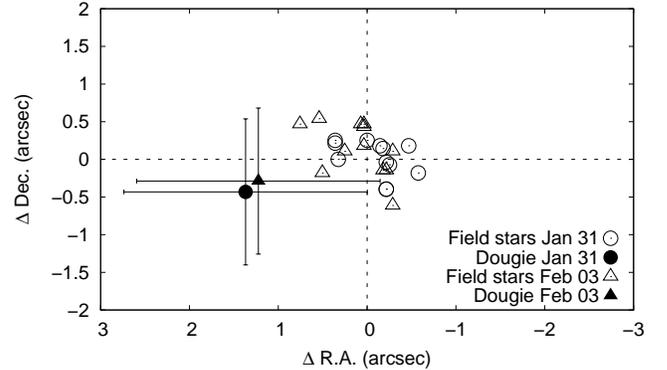}
\caption{The position of the transient (filled symbols) as measured on the
Jan 31 (circle) and Feb 03 (square) {\it Swift} U-band frames relative to the photometric
centroid of the host galaxy on the $r$-band SDSS frame. The error bars
represent 1 pixel uncertainty of the centroid on the {\it Swift} frames (see text). 
Open circles and triangles denote the position differences for the reference objects
on the same {\it Swift}- and SDSS frames. The scattering of the open symbols
around the center illustrate the registration uncertainty between the 
{\it Swift} and SDSS World Coordinate Systems. The transient appears $1.3$ arcsec 
($3.9$ kpc) off the photo-center of the host.}
\label{fig:pos}
\end{center}
\end{figure}

\begin{figure*}
\begin{center}
\epsscale{1.1}
\plottwo{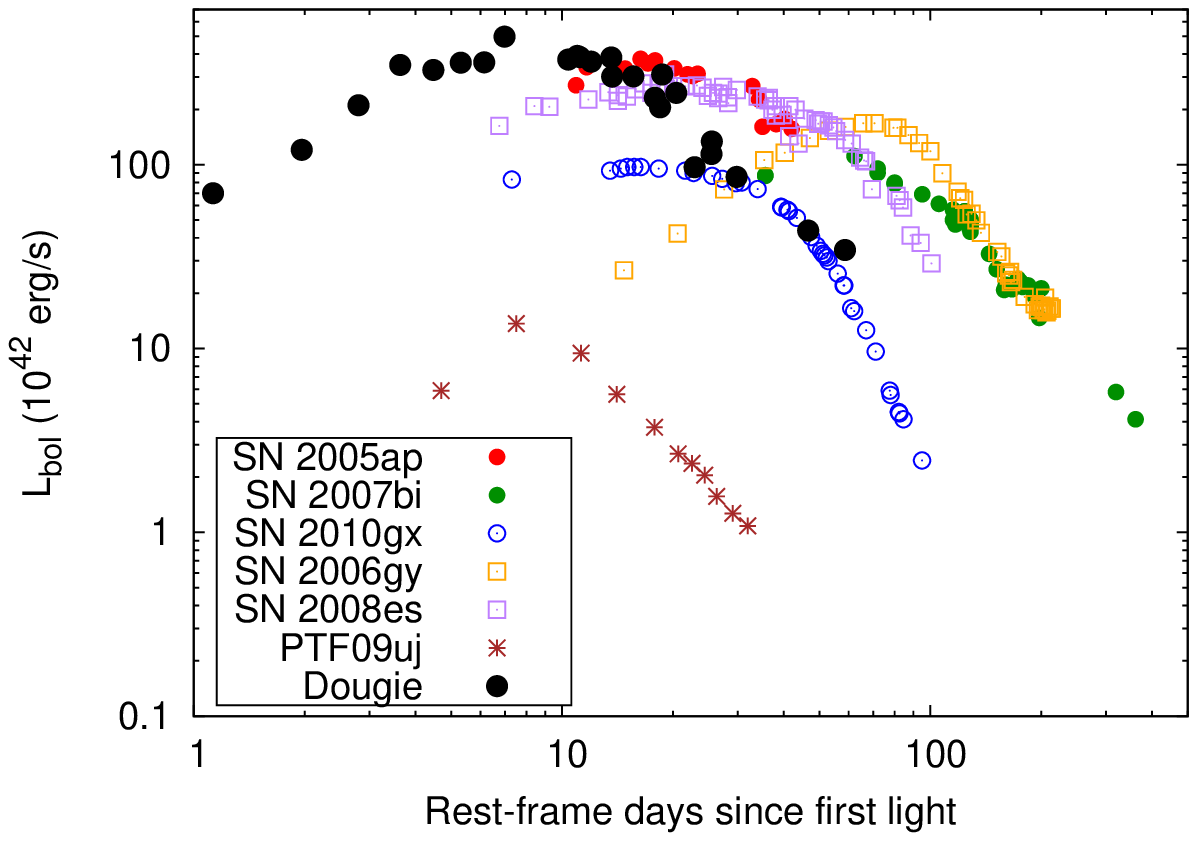}{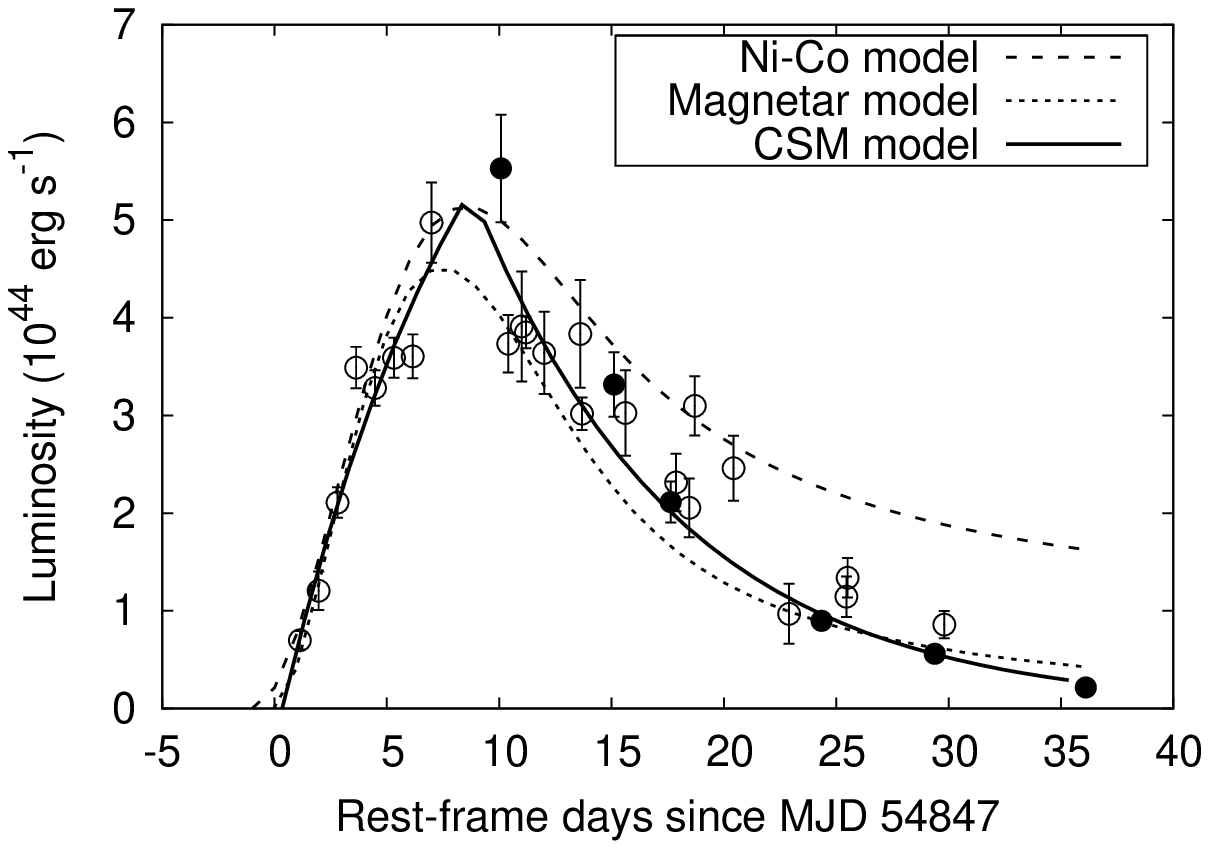}
\caption{Left panel: Comparison of the bolometric LC of {\it Dougie} with those of 
SLSNe. See \citet{manos11} for references to the SLSNe data. Right panel: three simple diffusion models constructed  to fit {\it Dougie}'s bolometric LC. 
{\it Filled} circles  correspond to data obtained from the SED-integration method described in Section~\ref{sec:obs}, while 
{\it open} circles denote bolometric fluxes estimated from photometry of both ROTSE and Swift detections. 
The {\it dashed}  curve shows  a model in which the  expanding SN ejecta is
heated by the radioactive decay of $^{56}\mathrm{Ni}$. The
{\it dotted} curve shows the LC from a magnetar-powered model, while the {\it solid} line depicts a  shocked, CSM  model.
The reader is refer to Section~\ref{sec:sn} for a brief discussion of the model parameters.}
\label{fig:bolom}
\end{center}
\end{figure*}

The bright, fast-rising transient appeared slightly offset from the 
centroid of the host galaxy as reported in the
SDSS DR10 catalog. Since the position of {\it Dougie} within
the host galaxy can be key for interpreting  its  physical origin,
here we investigate this issue in more detail.

In order to estimate the uncertainties associated with  {\it Dougie}'s position, we  first co-added {\it Swift}/UVOT frames obtained on Jan 31 and Feb 03
(+10 and +15 rest-frame days, respectively), when the OT was observed with 
the highest signal-to-noise.
We  then register the  {\it Swift} frames  to the SDSS r-band
frame of the same area by matching the positions of common point sources
on both frames.   Next, we determine the coordinates of {\it Dougie}  on the registered
{\it Swift} frames and compared them  to
those of the photometric centroid of the host galaxy  as measured 
on the SDSS r-band frame. The results are shown in Fig.~\ref{fig:pos},
where the {\it Swift} minus SDSS coordinate differences (in arcseconds) 
are plotted as $\Delta$ R.A. and $\Delta$ Dec. Open symbols
represent the coordinate differences for the reference objects, 
while the filled symbols denote the position of {\it Dougie} with respect to 
the host photo-center.

Fig.~\ref{fig:pos}  reveals that  that the reference objects scatter around  the zero point  more-or-less isotropically, 
and there is no obvious systematic shift between the data obtained on the epochs 
of the two {\it Swift} observations. On the other hand, {\it Dougie} appears to be systematically offset, about 1.2" East and 0.4" South
from the center of the host, in both observed epochs. Adopting
the WCS of the SDSS frame, we measure the final position of {\it Dougie} as
R.A.= $12^h 08^m 47.86^s$ ($\pm 0.02^s$), Dec.= $+43^o 01' 19.9"$ ($\pm 0.6"$).
We note that the  uncertainties  for {\it Dougie}  plotted in Fig.~\ref{fig:pos} are conservative
$\pm 1$-pixel errors of the {\it Swift} frames that may overestimate 
the true errors of the photo-centers. The coordinate uncertainties given above
are in between the $\pm 1$-pixel errors and the measured frame-to-frame shift 
of the photometric center of the OT ($\sim 0.1$ pixel) on the {\it Swift} frames.

Based on our {\it Swift} observations,  we infer that {\it Dougie} 
appears to be slightly  off-center from its host galaxy. The total angular distance
from the photo-center of the host is $\approx 1.3$ arcsec, corresponding to
$\approx  3.9$ kpc linear distance for the adopted redshift ($z = 0.191$, \S 2.4). 


\subsection{The quasi-bolometric light curve}

The bolometric lightcurve (LC)  of the transient was assembled by integrating the host-corrected 
UV+optical SEDs (Fig.~\ref{fig:sed}) over wavelength. The  fluxes
in the far-UV were estimated by linearly extrapolating the spectrum  until 
1000 \AA. In the IR (longward of 8000 \AA), a Rayleigh-Jeans tail  starting from
the reddest observed flux was assumed  to calculate the integrated  IR contribution to the total quasi-bolometric flux. 

The temporal coverage of the  LC was  refined
by adding more data points from the ROTSE $R$-band and Swift $v$-band LCs 
(Fig.~\ref{fig:lc}). This was justified by the close similarity between the
shape of the LCs at optical wavelengths. 
To do this, the observed ROTSE- and Swift magnitudes were converted to
absolute magnitudes using a $D_L = 897$ Mpc luminosity distance, 
then the luminosity curve was calculated from these absolute magnitudes as if they
were bolometric magnitudes, which is equivalent of assuming BC = 0 mag bolometric
correction. Although this seems like  a rather crude approximation, the 
resulting LC (plotted in Fig.~\ref{fig:bolom}) 
looks consistent with those  obtained from SED integration. The peak bolometric luminosity of {\it Dougie}  is thus estimated to be 
$L_{\rm peak} \approx 5 (\pm 1) \times 10^{44}$ erg s$^{-1}$, while the integrated  radiated energy  is calculated to be
$E_{\rm rad} \approx 6.2 (\pm 0.2) \times10^{50}$ erg.

\section{On the Origin of {\it Dougie}}

In this section we consider four  models for the origin of {\it Dougie}: 
core collapse supernova, merging neutron stars, off-axis GRB models and the tidal disruption of a star by
the central supermassive black hole.  These are presented in detail in the following subsections.

\subsection{Core Collapse  Models}\label{sec:sn}

Although the lack of SN features in the optical spectra does not support the core collapse hypothesis, the LC look rather similar to those of SLSNe.
For this reason, in the absence of spectral information one could have naturally deduced that {\it Dougie} might have resulted from the death of a  massive star that was recently formed  in the host galaxy.

In the left panel of Fig.~\ref{fig:bolom} we compare the bolometric LCs of {\it Dougie} with
those of several SLSNe. The data of the latter objects were analyzed by \cite{manos11} where the
reader may find the references to the data. It is seen that {\it Dougie} showed faster LC
evolution than most of the well-observed SLSNe. The rise-time to peak, $t_{rise} \sim 10$ days,
was similar to that of PTF09uj, a luminous Type IIn SN \citep{ofek10}. However, as Fig.~\ref{fig:bolom}
shows, the peak luminosity of {\it Dougie} clearly puts it into the SLSN regime.

In this subsection we make an effort to describe the LC using simple SN radiative 
diffusion models to ascertain whether or not they can provide a reasonable description 
of the bolometric LC.
Following the formalism developed  by \citet{manos11}, we test three different core collapse scenarios: a   {\it Ni-Co} radioactive decay model, a  magnetar model
\citep{kb10, woos10} and a  shocked, circumstellar medium (CSM)  model. The best-fit representations of the LC are plotted in Fig.~\ref{fig:bolom}.
All three models assume that the energy is deposited
at the center of an optically-thick sphere, promptly thermalized, and 
then slowly transported  out by photon diffusion.  In the radioactive decay and 
magnetar modes we assume homologous expansion of the SN ejecta (which
is taken into account when solving the diffusion equation), 
while in the CSM  model we assume a fixed, opaque CSM cloud whose interior  is thermalized
at a designated time.

The fast rise and decay  of the observed LC can be fit only using a relatively short effective diffusion timescale, which corresponds to a low ejecta mass in all three scenarios. 
The high peak luminosity, on the other hand,   requires very large internal  energy  to be readily available.
In the radioactive decay model, in particular, the derived  diffusion timescale, $t_{\rm d} \approx 7-8$ days,
implies $M_{\rm ej} \approx 1M_\sun$ for $\kappa= 0.1$ cm$^2$~g$^{-1}$ and $v_{\rm sn} = 3\times 10^4$  km~s$^{-1}$, while the 
large peak luminosity demands a Ni-mass $\approx 15M_\odot$.  In the magnetar model, the early peak  requires a short ($\sim 5$ days) spin-down timescale which, in turn, requires a relatively  large magnetic field strength of about $4 \times  10^{14}$ G for an initial 10 ms spin period. Moreover,
the magnetar model also needs a relatively low-mass ejecta ($M_{\rm ej} \sim 1$ $M_\odot$), but a very extended initial radius of $R_0 \sim 10^{14}$ cm. 
In addition, this model requires the internal  energy to be injected near the edge of the very tenuous SN  ejecta to avoid significant adiabatic degradation. 

The CSM model  gives a more physically 
consistent picture with a required total shocked energy  of about $ 8 \times 10^{50}$ erg (compatible with the explosion energy of a typical SN) deposited within a CSM cloud of about  $\approx 2.6$ $M_\odot$. In this scenario {\it Dougie} resembles to PTF09uj \citep{ofek10, manos11}, i.e.
the luminosity is due to the conversion of shock kinetic energy into thermal energy within the
opaque, dense CSM. The higher luminosity of {\it Dougie} might be explained by the larger kinetic
energy and denser, more massive CSM than in the case of PTF09uj. 
However, the weakness of this hypothesis is that, unlike in PTF09uj, there is no indication for any shock-generated emission lines in the spectra of {\it Dougie}, which are ubiquitously observed in interacting Type IIn SNe. In principle, the lack of hydrogen and/or helium emission lines might be consistent with the presence of a H/He-free, O-rich CSM around the transient, but the lack of
{\it any} kind of spectral feature during the whole observed period makes this hypothesis less
feasible.

The other important observational constraint that argues against the core-collapse SLSN scenario
is the nature of the host galaxy (\S 2.5). There is growing evidence that 
H-free SLSNe tend to appear in metal-poor, dwarf galaxies that show intense
star formation rates and extremely strong emission lines \citep{neill11, lunnan14, leloud14}.
Since the host of {\it Dougie} appears to be dominated by older populations of stars 
without any sign of enhanced star formation, these observed properties of the host galaxy strongly
argue against the SLSN nature of {\it Dougie}.

\subsection{Neutron star merger model}

Merging neutron stars (NS-NS mergers, or ``merger-novae'') are thought to be promising candidates for producing fast, luminous transients \citep{lee07, metzger10, roberts11, barnes13, yu13, metzger14}. Recently, the rapid optical transient
PTF11agg \citep{cenko13} was proposed to be due to such a phenomenon \citep{wu14}. In this model
the merging of two neutron stars due to gravitational wave radiation is speculated to produce, 
in some cases, a rapidly spinning, hypermassive, magnetized neutron star \citep{rosswog03} 
surrounded by a more-or-less spherical, fast-expanding ($v \sim 0.1c$)
envelope, the mass of which is $M_{\rm ej} \lesssim 0.1$ $M_\odot$. 
The magnetar wind is assumed to efficiently dissipate Poynting flux within the expanding envelope, heating it and accelerating it to relativistic speed ($v \sim c$). 
The  dilution of the envelope due to expansion causes the thermalized photons  to escape on timescales anywhere between hours and days depending primarily on the mass of the surrounding envelope and its expansion velocity.
Thus, the basic physical configuration, to some extent, is similar to the
magnetar model considered in \S 3.1.  The main difference being the mass of the ejected envelope, 
which in the SN model is at least an order of magnitude larger.

While the stable, hypermassive neutron stars model could produce LCs with shapes that are qualitatively similar  to that of {\it Dougie},
i.e. rapid rise followed by slower decline, the calculations by e.g. \citet{yu13} and \citet{metzger14}
show rise times that are significantly shorter (typically between 
10 hours - 1 day) than the $\sim 10$ days observed rise time of {\it Dougie}. This is essentially
due to the smaller ejected mass in the NS-NS merger systems and its corresponding higher expansion velocity. As it was shown in \S 3.1, the observed LC of {\it Dougie}
needs $t_{\rm d} \sim 8$ days, which is almost a factor of 10 longer than the typical
diffusion timescales expected in NS-NS merger systems. Therefore, we conclude that the predictions 
of the merger model are not compatible with the observed LC of {\it Dougie}.

\subsection{Orphan afterglow model}

\begin{figure}
\begin{center}
\epsscale{1.2}
\plotone{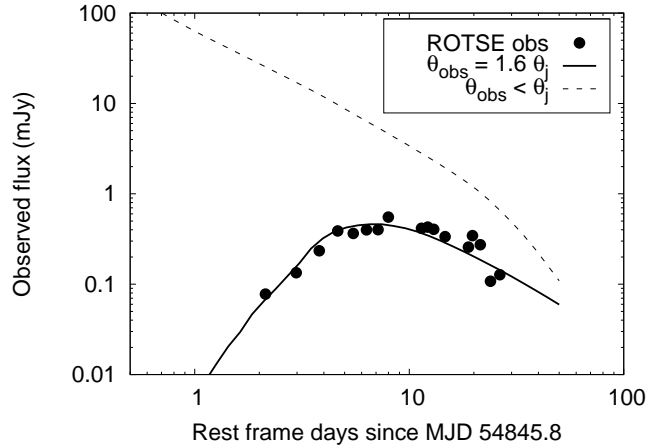}
\caption{Afterglow emission from an off-axis GRB jet. Light curves calculated for two viewing angles $\theta_{\rm obs}\leq \theta_{\rm j}$ (dashed curve) and $\theta_{\rm obs}=1.6 \theta_{\rm j}$ (solid curve), and  for a GRB with $\theta_{\rm j} = 0.3$,  $E_\Omega = 10^{54}$ erg,  $p = 2.5$, $n_0 = 1.0$ cm$^{-1}$, 
$\epsilon_B = 0.01$ and $\epsilon_e = 0.1$. The model parameters were chosen in order  to find an
acceptable match to the ROTSE unfiltered magnitudes ({\it filled} symbols), which correspond to  $\nu \approx  4.5 \times 10^{14}$ Hz. }
\label{fig:oag}
\end{center}
\end{figure}

\begin{figure}
\begin{center}
\epsscale{1.2}
\plotone{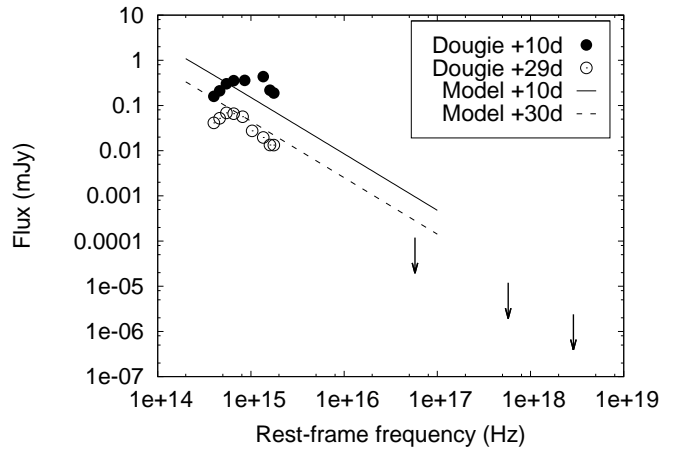}
\caption{Predicted synchrotron spectra of the $\theta_{\rm obs} =1.6 \theta_{\rm j} $ orphan
afterglow model in Fig.~\ref{fig:oag} compared with the observed optical-UV SEDs ({\it filled} symbols).
The X-ray upper limits calculated from the integrated {\it Swift}/XRT observations are indicated by 
the downward arrows. }
\label{fig:oag2}
\end{center}
\end{figure}

Given that most gamma-ray bursts (GRBs) are collimated into narrow jets, their observed properties will unavoidably  vary depending on the angle $\theta_{\rm obs}$ from their symmetry axis at which they are observed.  If {\it Dougie} were a GRB, then  at least its gamma-ray emission directed at us was certainly extremely weak. A plausible  interpretation might be that the 
{\it Dougie}  was a typical  GRB seen at an angle, $\theta_{\rm obs}$, larger than the opening angle of the central jet, $\theta_{\rm j}$.

If we assume a homogeneous sharp-edged jet, the burst seen by all observers located within the initial jet aperture, $\theta_{\rm obs} < \theta_{\rm j}$, is practically the same, but beyond the edges of the jet the emission declines precipitously \citep{woods1999,granot02,rr2005}. When $\theta_{\rm obs} \gg \theta_{\rm j}$, there is no detectable prompt  emission and the accompanying  early afterglow is weak, owing to  relativistic beaming of photons away from the line of sight. As the Lorentz factor decreases with time, an off-axis observer will see a rising afterglow light curve at early times peaking when the jet Lorentz factor reaches $\approx 1/(\theta_{\rm obs} - \theta_{\rm j})$ and approaching that seen by an on-axis observer at later times. This is because an observer will receive most emission from those portions of a GRB blast wave that are within an angle 1/$\Gamma$ to the direction to the line of sight such that the emission for an off-axis observer will remain at a very low level until the Doppler cone of the beam intersects the observer's line of sight. This can be seen by comparing the $\theta_{\rm obs} < \theta_{\rm j}$ and $\theta_{\rm obs} = 1.6 \theta_{\rm j}$ light curves in Fig.~\ref{fig:oag}.

The off-axis  GRB interpretation of {\it Dougie} requires the viewing angle to have been $\theta_{\rm obs} \approx  1.6\theta_{\rm j}$ (Fig.~\ref{fig:oag}), similar to
the parameters determined by \citet{cenko13} for PTF11-agg. The afterglow light curves  at  $\nu \approx  4.5 \times 10^{14}$ Hz presented here are calculated using the afterglow models  of \citet{vaneerten12} by applying the BOXFIT\footnote{\tt \small http://cosmo.nyu.edu/afterglowlibrary/boxfit2011.html} code.   In these models, the expansion of the GRB jet in a uniform  medium with density $n_0$  is calculated using relativistic hydrodynamical simulations \citep[e.g.][]{zhang2009,fabio2012a,vaneerten2010,fabio2012b}. The local emissivity is computed using the conventional assumptions of synchrotron emission from relativistic electrons that are accelerated behind the shock into a power-law distribution ($\propto \gamma^{-p}$), where the electrons and the magnetic field hold fractions $\epsilon_e$ and $\epsilon_B$, respectively, of the internal energy of the shocked fluid \citep{sari98}.  

One question that naturally arises is whether the observed multi-wavelength  evolution can be explained within the framework of this model. Even though the model fits the optical  light curves moderately well, it is inconsistent with the observed SED and, in addition, predicts significantly higher X-ray fluxes when compared with the {\it Swift}/XRT upper limits. This is illustrated in Fig.~\ref{fig:oag2}, where the predicted synchrotron spectra
(solid lines) are compared with the observed SED. This argues against an off-axis GRB origin  for {\it Dougie}.  

\subsection{TDE models}

\begin{figure}
\begin{center}
\epsscale{1.2}
\plotone{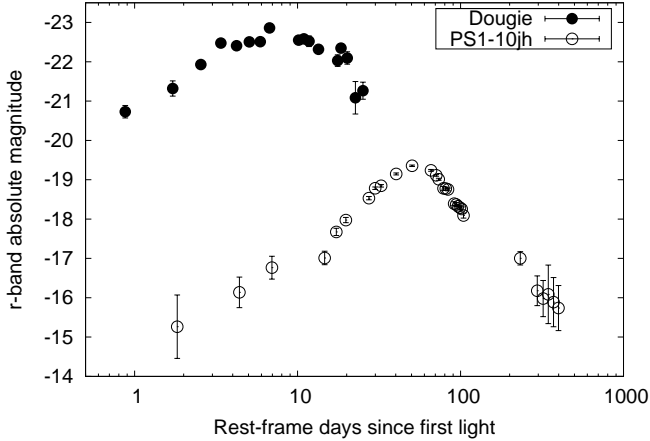}
\caption{$r$-band absolute light curves of {\it Dougie} (filled circles) and PS1-10jh (open circles),
another TDE-candidate \citep{gezari12}. For the latter, the redshift of $z=0.1696$ was applied to
correct for time dilation and distance.}
\label{fig:tdelc}
\end{center}
\end{figure}

The tidal disruption of an approaching star by a supermassive black hole (SMBH) has become
a popular mechanism for explaining the bright, slowly evolving
X-ray/UV/optical ``flares'' from luminous transients \citep[see e.g.][and
references therein]{chorn13, arcavi14, holoi14}. TDEs are characterized by the following basic quantities: SMBH mass $M_{\rm h}$, stellar mass $M_{\ast}$, stellar radius $R_{\ast}$, and the impact parameter, the ratio of the tidal 
radius $r_{\rm t} \equiv R_{\rm s} (M_{\rm h}/M_{\ast})^{1/3}$ to the distance of the closest approach $r_{\rm p}$,
as $\beta = r_t / r_p$ \citep{rees88}.

When the debris from a disrupted star falls back to
the SMBH, it first forms an elliptical accretion disk from which matter circularizes via a combination of mechanisms \citep{rr2009,hayasaki13,dai13,guill14}. The temporal evolution of the fallback rate $\dot{M}$ is thought to be characterized by a range of power law values $\sim t^{-n}$, where $n$ depends on the star's structure and the impact parameter $\beta$ \citep{lodato09, guill13}. The conversion of the fallback into light is not expected to directly follow $\dot{M}$, especially for individual bands \citep{strubbe09,lodato11}. As gas accretes onto the black hole, radiation pressure may expel some fraction of the mass if the accretion rate exceeds the Eddington limit of the black hole, $L_{\rm Edd} = 4 G M_{\rm h} m_{\rm p} c / \sigma_{\rm T}$. When this condition is satisfied, a wind may be produced that carries a significant amount of kinetic and thermal energy \citep{strubbe09, lodato11}. For cases in which the accretion rate remains sub-Eddington, the disk component likely becomes the dominant source of radiation, peaking in the far UV to soft X-ray bands; however, previous TDE candidates have shown evidence of significant reprocessing of the emergent light into longer wavelengths with an SED characterized by a single blackbody, rather than the sum of blackbodies expected for an accretion disk \citep{gezari12,guill14}.

Even in the observed bands, which do not include the peak in the SED at early times (see Fig.~\ref{fig:sp}), the peak luminosity of Dougie, $L_{\rm peak} \gtrsim 5 \times 10^{44}$ ergs s$^{-1}$, is larger than the Eddington limit of a $10^{7} M_{\odot}$ black hole. Additionally, Dougie's time of peak is significantly shorter than the time of peak predicted from $\dot{M}$ alone, which for a $10^{6} M_{\odot}$ black hole is on the order of a month for main-sequence stars \citep{guill13}. Both of these aspects suggest that if {\it Dougie} is indeed a tidal disruption event, its observational appearance near peak must be dominated by a wind component whose functional form may not directly reflect $\dot{M}$.

To model {\it Dougie}, we modified the {\tt TDEFit} code initially described in \citet{guill14} to include a wind component that can carry a significant fraction of the accretion energy. The appearance of a TDE when its accretion rate exceeds Eddington has been explored by a number of authors \citep{strubbe09,lodato11,coughlin14}. For super-Eddington accretion rates, these models presume a fraction of the incoming mass is ejected in the form of a wind, which moves out at some velocity that is comparable to the orbital velocity at the tidal disruption radius. If the majority of the energy carried by the wind is internal, the radiative output can be significantly less than the energy input, as much of the energy will be expended as work as the wind expands \citep{strubbe09}. If the excess energy is instead mostly carried kinetically, but then dissipates near the photosphere, the radiative output can be comparable to the energy input.

While these works provide descriptions of the relevant wind physics in the decline phase, they do not provide formalisms that are general enough to model {\it Dougie} over its full evolution: rise, peak, and decline. Additionally, each model has a particular prediction for the power-law relationship between $\dot{M}$ and the photosphere's properties. As an example, {\it Dougie} shows a clear decrease in temperature as a function of time after peak (Fig.~\ref{fig:sp}), whereas \citet{strubbe09} and \citet{lodato11} predict a temperature increase until the event drops below the Eddington limit. 
\citet{coughlin14}, by contrast, do predict a slight temperature decrease after peak.

\begin{figure}
\begin{center}
\epsscale{1.0}
\plotone{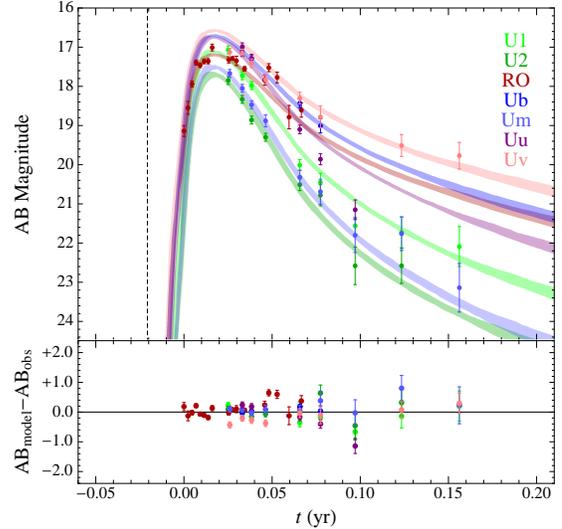}
\caption{Results of maximum-likelihood analysis performed by \tdefit. The figure shows the {\it Swift} and ROTSE-IIIb photometry as compared to the 1-$\sigma$ scatter of models with the highest likelihood. The top panel shows AB magnitudes of the data (dots) and models (shaded regions), while the bottom panel shows the difference between the data and the best-fit model.}
\label{fig:tde1}
\end{center}
\end{figure}

Fig.~\ref{fig:tdelc} shows the comparison of the r-band LC of {\it Dougie} with
that of another TDE-candidate, PS1-10jh \citep{gezari12}. It is seen that the two events had markedly
different light curves: {\it Dougie} showed a much faster and more energetic outburst
than PS1-10jh. 
Modeling of PS1-10jh has shown that a standard thin-disk alone is not capable of fitting the event's light curve \citep{guill14}, but that a large fraction of the disk's light needs to be intercepted by a reprocessing region to adequately match observations. 
We propose that the reprocessing region intercepts a fraction, $f_{rep}$, of the disk radiation. 
The reprocessing region is likely to be hydrostatic when the accretion rate is sub--Eddington, 
but to become dynamic and unbound when the Eddington limit is exceeded. The subsequent dynamic 
expansion then releases energy that originates from a combination of the radiation from the disk 
and radiation from the expanding reprocessing region,
\begin{equation}
L = (1 - f_{\rm rep}) L_{\rm disk} + L_{\rm rep}.
\end{equation}
We assume $L_{\rm disk}$ is capped at the Eddington luminosity, and $L_{\rm rep}$ is equal to the fraction of reprocessed disk light plus a fraction $f_{\rm out}$ of the Eddington excess,
\begin{equation}
L_{\rm rep} = f_{\rm rep} L_{\rm disk} + \eta f_{\rm out} \left(\dot{M} - \dot{M}_{\rm Edd}\right) c^{2},
\end{equation}
where we have presumed that the maximum amount of energy released in the form of an outflow is given by the energy release at the
innermost stable circular orbit (ISCO), $\eta c^{2}$, where the black hole efficiency $\eta$ depends solely on the black hole's spin parameter $a_{\rm spin}$. As in \citet{guill14}, we do not presume an {\sl a priori} time-dependence of the photosphere on $\dot{M}$, its optical depth $\tau$, its size $R_{\rm ph}$, or its temperature $T_{\rm ph}$, but rather leave these as free parameters. Because this model does not presume a particular power-law relationship between $\dot{M}$ and the reprocessing region's properities, the model space includes the specific power-law index proposed in \citet{strubbe09} and \citet{lodato11}, which would be favored by the Markov-chain Monte Carlo (MCMC) optimization if they are able to reproduce Dougie's observed evolution. One simplification made here is that we presume the power-law relationship between $\dot{M}$, $R_{\rm ph}$, and $T_{\rm ph}$ is constant throughout the event, regardless of whether the event is above or below the Eddington limit, whereas \citet{strubbe09} and \citet{lodato11} advocated a transition at the Eddington limit. We find that such a transition is not necessary to produce satisfactory fits (Fig.~\ref{fig:tde1}), but relaxing this assumption may improve fit quality.

Beside the parameters described above ($M_h$, $M_{\ast}$, $\beta$, $a_{\rm spin}$, $R_{\rm ph}$, $\tau$ and $f_{\rm out}$) 
the model also includes the following additional parameters \citep[see][for more complete description]{guill14}: 
the power-law index $l$ in the relationship $R_{\rm ph} \propto \dot{M}^l$; 
the disk inclination angle $\phi$ ($\phi = 0$ indicating face-on); the disk viscosity parameter $\V$;  
the polytropic index $\gamma$ of the disrupted star (either $5/3$ or $4/3$); the hydrogen column density $N_H$ within
the host galaxy; the reddening-law parameter $R_V$. We assumed that the time-lag between the disruption and
the first detection is $t_{\rm off}$ (in days), and we added $\sigma_v$ variance (in magnitudes) to the model light curves.

\subsubsection{Properties of highest-likelihood TDE models}

In Table~\ref{tab:tde} we show the median values resulting from our maximum-likelihood analysis. As expected, the short peak timescale and high luminosity of {\it Dougie} favors a low mass black hole, although there is very large scatter in the favored black hole mass ($\Log_{10} M_{\rm h} = 5.29_{-0.45}^{+0.91}$) so that the black hole mass might be as large as 
$\approx 10^6 M_{\odot}$. If we assume that the host galaxy's light is dominated by a bulge component, the black hole mass it would possess based on the \citet{haring04} relation would be $\approx 10^{7} M_{\odot}$; however as the galaxy is Sb type, its bulge fraction may be as low as $20\%$ \citep{binney98}, suggesting a central black hole mass of a few $10^{6} M_{\odot}$. 
This value is within the scatter of the black hole mass distribution found here. Another possibility is that the tidal disruption occurred about a secondary, smaller black hole in the process of merging with galaxy's primary black hole, which would also explain the TDE's slight offset from the host's center.

Lower mass ($10^5$ - $10^6$ $M_\odot$) black holes, even off-center ones, although rare, 
are not unprecedented. 
A good example is in NGC~3341, where recent merging 
resulted in two off-center nuclear sources (at $\sim 5$ and $\sim 8$ kpc from the center of
the massive disk galaxy), one of them being a Seyfert-2 nucleus \citep{barth08}.   
Another lower mass accreting BH ($\sim 10^6$ $M_\odot$) was recently discovered
in the central region of the dwarf starburst galaxy Henize 2-10 \citep{reines12}. The more recent
detection of a radio outburst from the ultra-luminous X-ray source HLX-1 in ESO 243-49 may be due
to a jet ejection from an intermediate-mass ($10^3$ - $10^4$ $M_\odot$) BH \citep{webb14}. 
As illustrated by these examples, the off-center position ($\sim 4$ kpc) of {\it Dougie}, as well
as the BH mass derived in the TDE-model, are not unrealistic, and may be consistent with the
TDE-hypothesis. 

\begin{figure}
\begin{center}
\epsscale{1.0}
\includegraphics[width=0.8\linewidth]{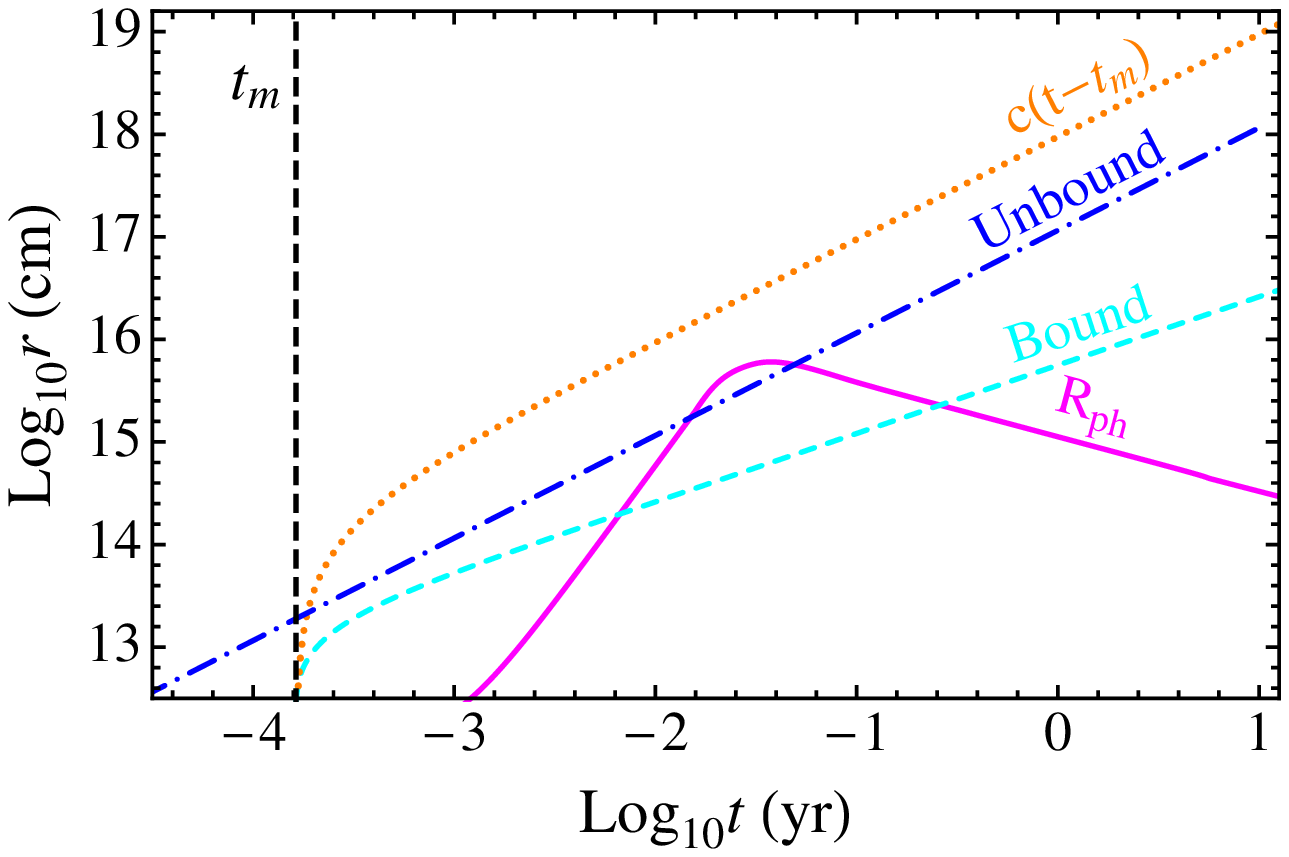}\\
\includegraphics[width=0.8\linewidth]{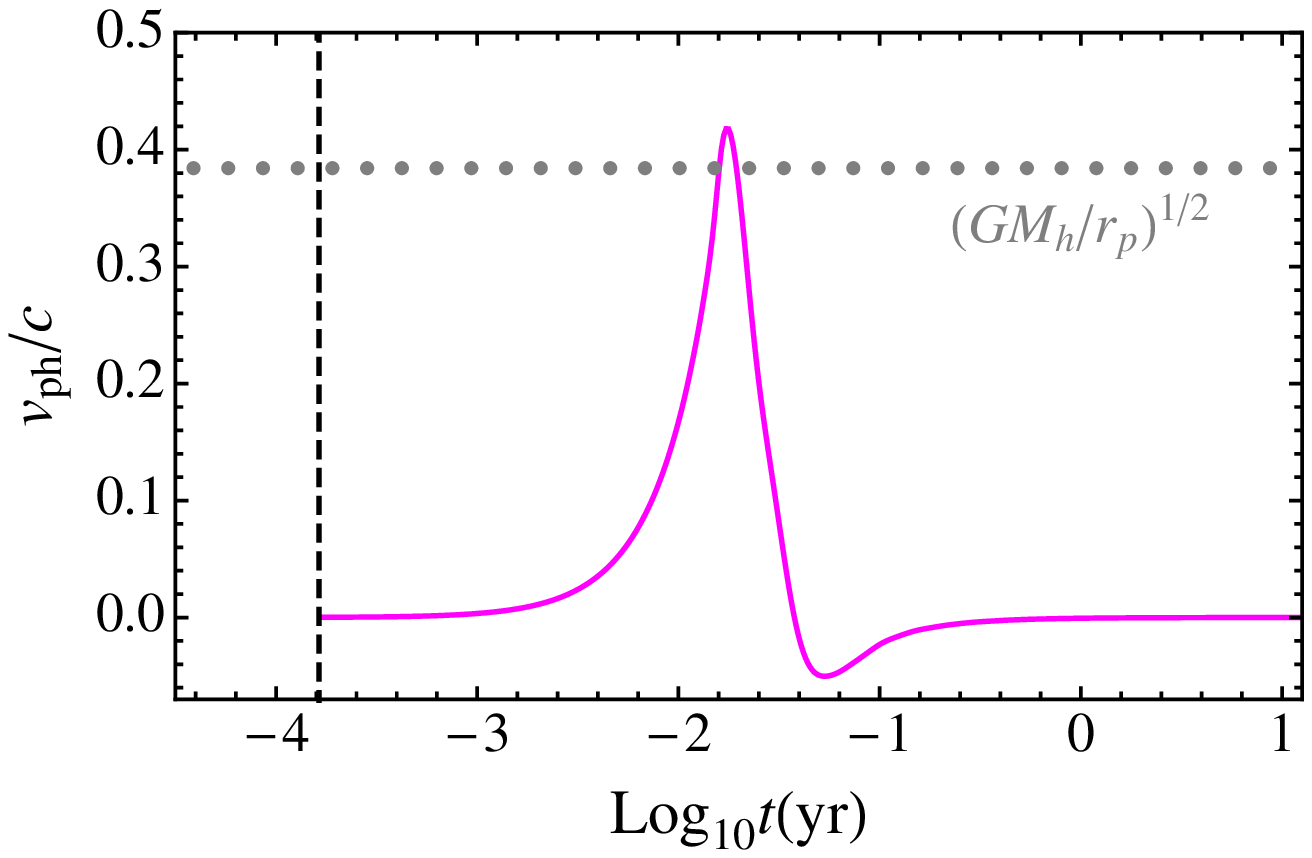}\\
\includegraphics[width=0.8\linewidth]{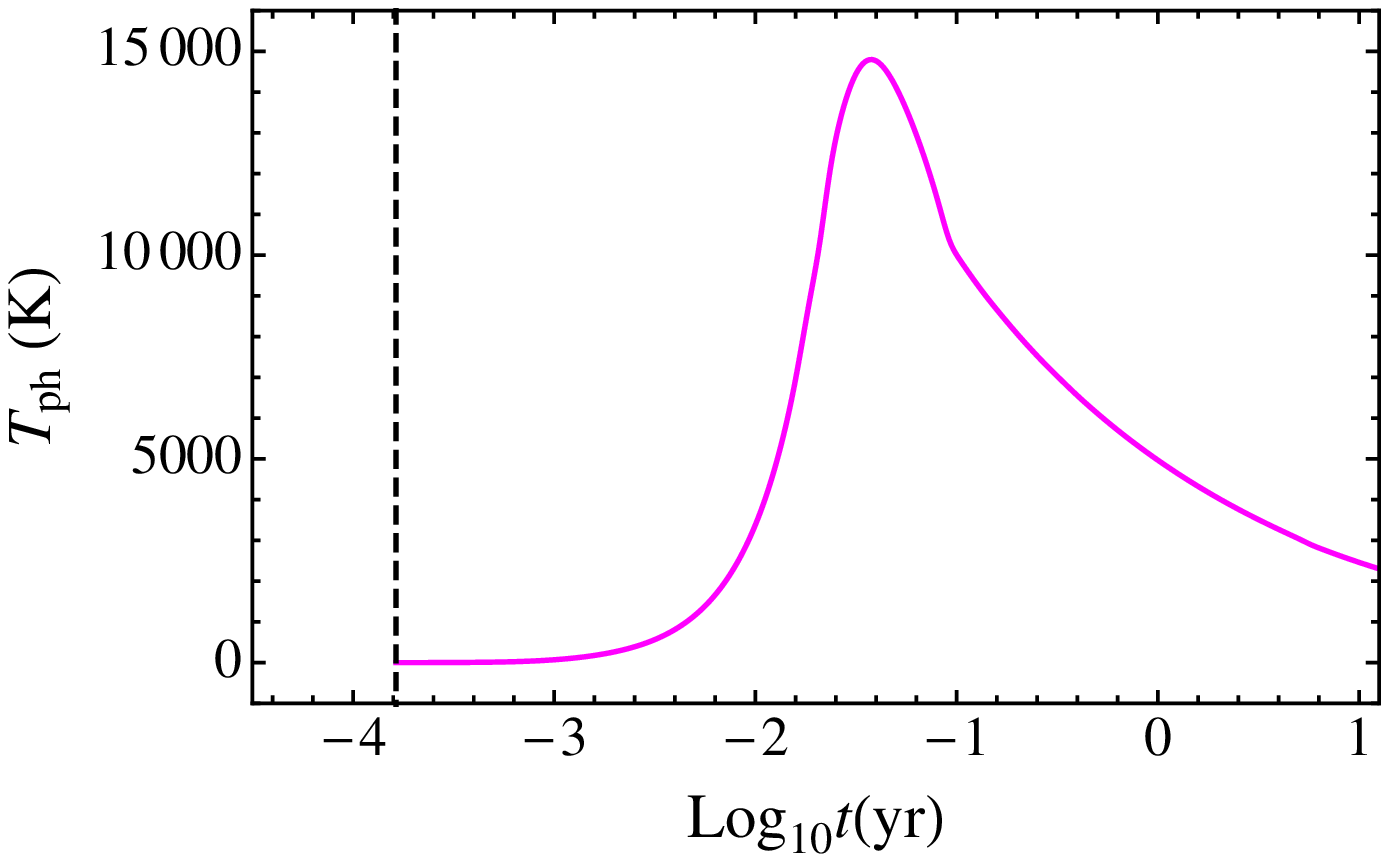}
\caption{Evolution of various quantities as functions of time for the TDE scenario for the highest-likelihood match to {\it Dougie}, where the vertical dashed times indicate the time of return of the most bound material $t_{\rm m}$. The top panel shows distance of various features as a function of time: the bound debris ({\it cyan, dashed}), the unbound tidal tail ({\it blue, dot-dashed}), the light travel distance since the time of disruption ({\it orange, dotted}), and the location of the wind/reprocessing photosphere ({\it magenta, solid}). The middle panel shows the velocity of the photosphere $v_{\rm ph}$, where the gray dotted line shows the escape velocity at twice the periapse distance. The bottom panel shows the temperature of the photosphere $T_{\rm ph}$.}
\label{fig:ph}
\end{center}
\end{figure}

A number of differences in {\sl Dougie}'s highest-likelihood parameters are apparent when comparing to fits of PS1-10jh. The best-fitting photosphere power-law index we find is small, $l = 0.16$. This parameter is related to the temperature power-law index by the expression $2l + 4m = 1$ \citep{guill14}. This implies that $T_{\rm ph} \propto \dot{M}^{0.17}$, close to the expectation of $\dot{M}^{1/4}$ for a blackbody the luminosity of which is proportional to $\dot{M}$ with a fixed $R_{\rm ph}$. Fig.~\ref{fig:ph} shows that the photosphere grows to a few $10^{15}$ cm (top panel), at velocity comparable to the escape velocity at $2 r_{\rm p}$ (middle panel), the terminal velocity for a wind launched from that location \citep{strubbe09}. The photosphere grows beyond the distance to which both the bound debris and unbound tidal tail have traveled since the time of disruption, suggesting that it must be continually accelerated; however, as the area of the photosphere increases more slowly than $\dot{M}$, $T_{\rm ph}$ also rapidly increases near peak and then cools off slowly at later times (Fig.~\ref{fig:ph}, bottom panel). For PS1-10jh, it was found that the reprocessing region's temperature was inversely related to $\dot{M}$ \citep{guill14}.

\begin{table}
\begin{center}
\caption{\label{tab:tde}Parameters of Highest Likelihood TDE Models}
\begin{tabular}{cccll}
\hline
\hline
\colhead{Parameter\tablenotemark{a}} & \colhead{Units} & \colhead{Prior} & \colhead{Allowed Range} & \colhead{Value\tablenotemark{b}}
\\
\hline
\\[-1.1em]
$t_{\rm off}$ 				& days 				& Flat & 
\phm{`}$-10^{3} \rsep 10^{3}$ & \phs\phn\phn$7.6_{-1.0}^{+1.3}$ \\[2 pt]
$\Log_{10} M_{\ast}$ 		& $M_{\odot}$ 		& Kroupa & 
\phd\phn\phn$-3 \rsep 2$ & $-0.098_{-0.484}^{+0.449}$ \\[2 pt]
$\gamma$ 			& \nodata 			& Flat & 
\phs\phd\phn\phn4/3 or 5/3 & \phn\phn5/3 (99.7\%) \\[2 pt]
$\Log_{10} M_{\rm h}$ 		& $M_{\odot}$ 		& Flat & 
\phs\phd\phn\phn$4 \rsep 8.6$ & \phs\phn$5.29_{-0.45}^{+0.91}$ \\[2 pt]
$\beta$ 						& \nodata 			& $\beta^{-2}$ & 
\phs\phn$0.5 \rsep 4$ & \phs\phn$0.97_{-0.15}^{+0.19}$ \\[2 pt]
$a_{\rm spin}$ 				& \nodata 			& Flat & 
\phs\phn\phn\phd$0 \rsep 0.998$ & \phs\phn$0.64_{-0.45}^{+0.29}$ \\[2 pt]
$\Log_{10} \V$ 				& \nodata 			& Flat & 
\phd\phn\phn$-4 \rsep 0$ & \phn$-0.19_{-0.19}^{+0.13}$ \\[2 pt]
$\phi$ 						& radians 			& Flat & 
\phs\phd\phn\phn$0 \rsep \pi/2$ & \phs\phn$0.51_{-0.36}^{+0.35}$ \\[2 pt]
$\Log_{10} \tau$ 			& \nodata			& Flat & 
\phd\phn\phn$-6 \rsep 6$ & \phs\phn\phn$3.1_{-2.4}^{+2.1}$ \\[2 pt]
$l$ 							& \nodata 			& Flat & 
\phd\phs\phn\phn$0 \rsep 4$ & \phs\phn$0.16_{-0.06}^{+0.06}$ \\[2 pt]
$\Log_{10} R_{\rm ph, 0}$ 	& \nodata 			& Flat & 
\phd\phn\phn$-4 \rsep 4$ & \phs\phn$0.98_{-0.26}^{+0.11}$ \\[2 pt]
$\Log_{10} f_{\rm out}$ 			& \nodata 			& Flat & 
\phd\phn\phn$-4 \rsep 0$ & \phn$-0.16_{-0.29}^{+0.12}$ \\[2 pt]
$R_{\rm V}$ 				& \nodata 			& Flat & 
\phs\phd\phn\phn$2 \rsep 10$ & \phs\phn\phn$6.1_{-2.7}^{+2.7}$ \\[2 pt]
$\Log_{10} N_{\rm H}$ 		& cm$^{-2}$ 			& Flat & 
\phd\phs\phn$17 \rsep 23$ & \phs\phn\phn\phd$19_{-1.3}^{+1.4}$ \\[2 pt]
$\sigma_{\rm v}$ 			& \nodata 			& Flat & 
\phs\phd\phn\phn$0 \rsep 1$ & \phs\phn$0.24_{-0.04}^{+0.04}$ \\[2 pt]
\hline
\hline
\end{tabular}
\end{center}
$^{\rm a}$See \citet{guill14} for more detailed description.\\
$^{\rm b}$Median value, with ranges corresponding to 1-$\sigma$ spread from median.
\end{table}

\begin{figure}
\begin{center}
\epsscale{1.0}
\plotone{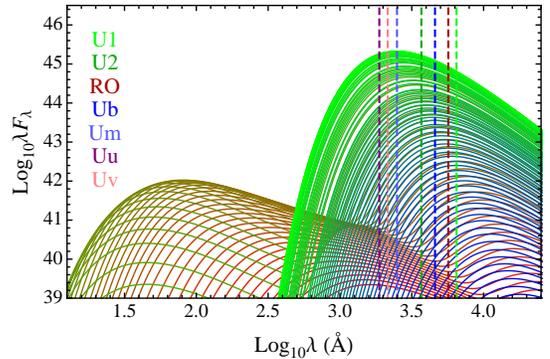}
\caption{Spectral energy distributions (SEDs) as a function of time for the highest-likelihood fit to {\it Dougie}. {\it Red} curves correspond to early times, {\it green} curves correspond to the flare at peak luminosity, and {\it blue} curves correspond to late times. The vertical {\it dashed} lines show the centroid of each filter used to observe {\it Dougie}, color-coded to match the filter designations in the top-left.}
\label{fig:tdesed}
\end{center}
\end{figure}

The models also favor nearly 100\% conversion of both the kinetic energy from the wind and radiative energy from the disk into energy radiated by the reprocessing photosphere (i.e. $f_{\rm out} \sim 1$ and $\tau \rightarrow \infty$), suggesting that the wind component is completely dominant. This conclusion is bolstered by the strong upper limits in the X-rays from Swift (see Section \ref{sec:xray}) that suggest $<1$\% of the radiative output emerges with energies above 200 eV. Our highest likelihood models suppress the total X-ray output to $< 10^{42}$ erg s$^{-1}$ (Fig.~\ref{fig:tdesed}). By contrast, the reprocessing zone in PS1-10jh was found to only intercept $\sim 1/3$ of the disk's radiative output. To see if such a high conversion factor was necessary, we performed a test \tdefit run in which $f_{\rm out}$ was fixed to 0.1, this yielded a poor fit and tended to even lower black hole masses ($M_{\rm h} < 10^{5}$). The low X-ray flux also suggests that if a jet were produced \citep[e.g][]{decolle2012}, it was at the very least not pointed towards Earth, and the high conversion ratio into the reprocessing zone may be the result of the jet being intercepted by a thick, super-Eddington accretion flow \citep{tchekhovskoy14}.

Aside from these differences, the favored stellar mass ($M_{\ast} = 0.8 M_{\odot}$), impact parameter ($\beta = 0.97$, indicating a full disruption for the favored $\gamma$), black hole spin ($a = 0.64$), and viscous parameter ($\V = 0.65$) are all typical values expected for a main-sequence disruption, suggesting that this event would be representative of disruptions about lower-mass black holes if it is in fact a tidal disruption.

\section{Summary}

As a summary, we draw the following conclusions:

\begin{enumerate}

\item{We discovered an optical transient (nicknamed {\it Dougie}) which was most probably of extragalactic origin.
It appeared on top of a faint galaxy having $z = 0.191$ redshift corresponding to
$D \approx 900$ Mpc distance. Our follow-up observations in the optical and UV-bands revealed that
the light curve of the transient showed a quick rise (with rise time of $\sim 10$ days) followed
by a subsequent decline resulting in a $\approx 1$ month-long observability with our resources.
The observed LC suggested a peak luminosity of $\approx 5 \times 10^{44}$ erg s$^{-1}$, which is
similar to those of the most luminous SLSNe.  
Simultaneous X-ray observations with {\it Swift} resulted in no detected X-ray emission on the
co-added {\it XRT} frames spanning the whole duration of the follow-up observations. }

\item{Our spectroscopic follow-up observations showed that the spectra of the transient 
were unusual: between 4000 and 9000 \AA\ it did not show any spectral feature
that could be attributed to the transient, during the whole observed interval. The weak 
narrow features in the late-time spectra were identified as due to the host galaxy. 
At early times the spectra consisted of a hot, blue continuum ($T \approx 16,000$ K), which
later cooled down, but could not be described as a single blackbody from the UV to the
red. The softening of the spectra suggested a cooling, expanding, SN-like ejecta, but
the lack of spectral features during the entire observed time domain argued against 
the SN-hypothesis.} 

\item{The proposed host galaxy, SDSS J120847.77+4320.1, is a faint Sb-type galaxy,
without any previously detected UV/X-ray emission, and without any sign of ongoing
star formation. The transient appeared  $\approx 1.3$ arcsec off-center, 
corresponding $\approx 3.9$ kpc linear distance from the photo-center of the host (2$\sigma$).}

\item{Despite the similar peak luminosity, it is unlikely that the transient was a super-luminous
supernova. Traditional SN models based on radioactive decay are ruled out because
of the order-of-magnitude difference between the required amount of  $^{56}$Ni 
mass ($\approx 15$ $M_\odot$) and ejecta mass ($\approx 1$ $M_\odot$, from LC rise time). 
The magnetar-powered, and the CSM-interaction-powered SLSN models can be tweaked 
to produce a similar LC, but the  difference between the usual spectra of 
these kind of transients and that of {\it Dougie} does not support the SLSN hypothesis. 
Also, the lack of ongoing star formation within the host galaxy is not typical for
galaxies producing H-poor SLSNe, as those SLSNe tend to appear in metal-poor hosts 
showing intense star formation.}

\item{Merging neutron stars that can produce fast, luminous transients, 
where the remnant collapse is halted when a rapidly rotating, hypermassive magnetar 
is produced, usually result
in light curves that evolve much faster than {\it Dougie}. Their predicted rise times, 
$t_{\rm rise} \sim $1-2 days \citep{metzger14}, are in contrast to the
observed $\sim 10$ days rise time of {\it Dougie}.}

\item{Based on the similarity of the hot featureless spectra to the early spectra of
GRB-SNe, afterglow models by \citet{vaneerten12} were fit to the LC and SEDs 
of {\it Dougie}. Both the shape and the peak of the LC could be explained by 
a jet-induced afterglow having parameters more-or-less similar to those derived by
\citet{cenko13} for PTF11-agg. All such models, however, fail to reproduce the 
observed SED evolution. }

\item{The tidal disruption scenario was explored by fitting the event to an amended version of the model presented in \citet{guill14}. The TDE model yielded a good fit to the photometric and spectral evolution of the flare, with the highest-likelihood models suggesting a disruption of a solar-mass star by a black hole. The BH mass turned out to be on the low-end of predictions for the associated host galaxy. Due to the slight offset of the flare from the host's center, this may be attributable to a recent merger of a lower-mass galaxy with the host. The TDE model also finds that the flare must have been very super-Eddington at peak with a near full conversion of energy released at the ISCO into energy radiated at its photosphere. As no X-rays were detected from the flare, there is no direct evidence for the existence of a jet, suggesting that the jet was ``smothered'' by the super-Eddington accretion flow, which is consistent with the near-full conversion of accretion energy into radiative output. The other parameters of the disruption yielded by our analysis are as expected of a typical disruption. We thus conclude that
{\it Dougie} could represent a canonical TDE about lower-mass central black holes.}

\end{enumerate}

\acknowledgements
We thank the anonymous referee for the thorough report that was helpful while revising the
first version of this paper.
This work has been supported by NSF Grant AST 11-09881 (UT, PI Wheeler), Hungarian OTKA Grant NN-107637 (Szeged, PI Vinko),  
NSF grant AST-0847563 (UCSC, PI Ramirez-Ruiz), and the David and Lucile Packard Foundation (ERR). GHM is supported by
NSF Grant AST 09-07903 (CfA, PI Kirshner). J.G. is supported by Einstein grant PF3-140108. 
E.C. was supported by the University of Texas at Austin Graduate Scholl Power's Fellowship, 
and is currently supported by the Enrico Fermi Institute via the Enrico Fermi Postdoctoral Fellowship.
ROTSE-III has been supported by NASA grant NNX-08AV63G, NSF Grant PhY-0801007, 
the Australian Research Council, the University of New South Wales, 
the University of Texas and the University of Michigan. 
The Marcario Low Resolution Spectrograph is named for Mike Marcario of High Lonesome Optics who fabricated several optics for the instrument but died before its completion. The LRS is a joint project of the Hobby-Eberly Telescope partnership and the Instituto de Astronom\'ia de la Universidad Nacional Aut\'onoma de M\'exico.
We acknowledge the HET Resident Astronomer Team (M. Shetrone, S. Odewahn, J. Caldwell, 
S. Rostopchin) for their work and support during the spectroscopic follow-up
observations. The NASA ADS and NED services were extensively used during the entire project, and the availability
of these services are also gratefully acknowledged.

\end{document}